\documentclass[useAMS,usenatbib]{mn2e}
\usepackage{rotating}
\usepackage{amssymb} 
\usepackage{amsmath} 
\usepackage{graphicx}
\usepackage{multirow}

\everymath{\displaystyle}

\author[Noutsos et al.]
{A.~Noutsos$^1$, D.~H.~F.~M.~Schnitzeler$^1$, E.~F.~Keane$^1$, M.~Kramer$^1$ and S.~Johnston$^2$\\
$^1$ Max-Planck-Institut f\"ur Radioastronomie, Auf dem H\"ugel 69, 53121 Bonn, Germany.\\
$^2$ Australia Telescope National Facility, CSIRO, P.O. Box 76, 
Epping, NSW 1710, Australia.\\
} 
\date{\today} 
\title[Pulsar Spin--Velocity Alignment II]
{Pulsar Spin--Velocity Alignment: Kinematic Ages, Birth Periods and Braking Indices}

\def\jnl@style=\rm
\def\ref@jnl#1{{\jnl@style#1}}

\def\aj{\ref@jnl{AJ}}                   
\def\actaa{\ref@jnl{Acta Astron.}}      
\def\araa{\ref@jnl{ARA\&A}}             
\def\apj{\ref@jnl{ApJ}}                 
\def\apjl{\ref@jnl{ApJ}}                
\def\apjs{\ref@jnl{ApJS}}               
\def\ao{\ref@jnl{Appl.~Opt.}}           
\def\apss{\ref@jnl{Ap\&SS}}             
\def\aap{\ref@jnl{A\&A}}                
\def\aapr{\ref@jnl{A\&A~Rev.}}          
\def\aaps{\ref@jnl{A\&AS}}              
\def\azh{\ref@jnl{AZh}}                 
\def\baas{\ref@jnl{BAAS}}               
\def\bac{\ref@jnl{Bull. astr. Inst. Czechosl.}}
\def\caa{\ref@jnl{Chinese Astron. Astrophys.}}
\def\cjaa{\ref@jnl{Chinese J. Astron. Astrophys.}}
\def\icarus{\ref@jnl{Icarus}}           
\def\jcap{\ref@jnl{J. Cosmology Astropart. Phys.}}
\def\jrasc{\ref@jnl{JRASC}}             
\def\memras{\ref@jnl{MmRAS}}            
\def\mnras{\ref@jnl{MNRAS}}             
\def\na{\ref@jnl{New A}}                
\def\nar{\ref@jnl{New A Rev.}}          
\def\pra{\ref@jnl{Phys.~Rev.~A}}        
\def\prb{\ref@jnl{Phys.~Rev.~B}}        
\def\prc{\ref@jnl{Phys.~Rev.~C}}        
\def\prd{\ref@jnl{Phys.~Rev.~D}}        
\def\pre{\ref@jnl{Phys.~Rev.~E}}        
\def\prl{\ref@jnl{Phys.~Rev.~Lett.}}    
\def\pasa{\ref@jnl{PASA}}               
\def\pasp{\ref@jnl{PASP}}               
\def\pasj{\ref@jnl{PASJ}}               
\def\rmxaa{\ref@jnl{Rev. Mexicana Astron. Astrofis.}}%
\def\qjras{\ref@jnl{QJRAS}}             
\def\skytel{\ref@jnl{S\&T}}             
\def\solphys{\ref@jnl{Sol.~Phys.}}      
\def\sovast{\ref@jnl{Soviet~Ast.}}      
\def\ssr{\ref@jnl{Space~Sci.~Rev.}}     
\def\zap{\ref@jnl{ZAp}}                 
\def\nat{\ref@jnl{Nature}}              
\def\iaucirc{\ref@jnl{IAU~Circ.}}       
\def\aplett{\ref@jnl{Astrophys.~Lett.}} 
\def\apspr{\ref@jnl{Astrophys.~Space~Phys.~Res.}}
\def\bain{\ref@jnl{Bull.~Astron.~Inst.~Netherlands}} 
\def\fcp{\ref@jnl{Fund.~Cosmic~Phys.}}  
\def\gca{\ref@jnl{Geochim.~Cosmochim.~Acta}}   
\def\grl{\ref@jnl{Geophys.~Res.~Lett.}} 
\def\jcp{\ref@jnl{J.~Chem.~Phys.}}      
\def\jgr{\ref@jnl{J.~Geophys.~Res.}}    
\def\jqsrt{\ref@jnl{J.~Quant.~Spec.~Radiat.~Transf.}}
\def\memsai{\ref@jnl{Mem.~Soc.~Astron.~Italiana}}
\def\nphysa{\ref@jnl{Nucl.~Phys.~A}}   
\def\physrep{\ref@jnl{Phys.~Rep.}}   
\def\physscr{\ref@jnl{Phys.~Scr}}   
\def\planss{\ref@jnl{Planet.~Space~Sci.}}   
\def\procspie{\ref@jnl{Proc.~SPIE}}   

\makeatletter
\DeclareRobustCommand{\Cpp}
{\valign{\vfil\hbox{##}\vfil\cr
   \textsf{C\kern-.1em}\cr
   $\hbox{\fontsize{\sf@size}{0}\textbf{+\kern-0.05em+}}$\cr}%
}
\makeatother

\begin{document}

\bibliographystyle{mn2e}

\maketitle

\begin{abstract}This paper presents a detailed investigation of the dependence of pulsar spin--velocity alignment, which has been observed for a sample of 58 pulsars, on pulsar age. At first, our study considers only pulsar characteristic ages, resulting in no change in the degree of correlation as a function of age, up to at least 100 Myr. Subsequently, we consider a more reliable estimate of pulsar age, the kinematic age, assuming that pulsars are born near the Galactic plane. We derive kinematic ages for 52 pulsars, based on the measured pulsar proper motions and positions, by modelling the trajectory of the pulsars in a Galactic potential. The sample of 52 pulsar kinematic ages constitutes the largest number of independently estimated pulsar ages to date. Using only the 33 most reliable kinematic ages from our simulations, we revisit the evolution of spin--velocity alignment, this time as a function of kinematic age. We find that the strong correlation seen in young pulsars is completely smeared out for pulsars with kinematic ages above 10 Myr, a length of time beyond which we expect the gravitational pull of the Galaxy to have a significant effect on the directions of pulsar velocities. In the discussion, we investigate the impact of large distance uncertainties on the reliability of the calculated kinematic ages. Furthermore, we present a detailed investigation of the implications of our revised pulsar ages for the braking-index and birth-period distributions. Finally, we discuss the predictions of various SN-kick mechanisms and their compatibility with our results.

\vspace{0.3cm}

\noindent {\bf Key words:} pulsars: general --- Galaxy: kinematics and dynamics --- methods: numerical.

\end{abstract}

\begin{figure*} 
\vspace*{10pt} 
\includegraphics[width=1\textwidth]{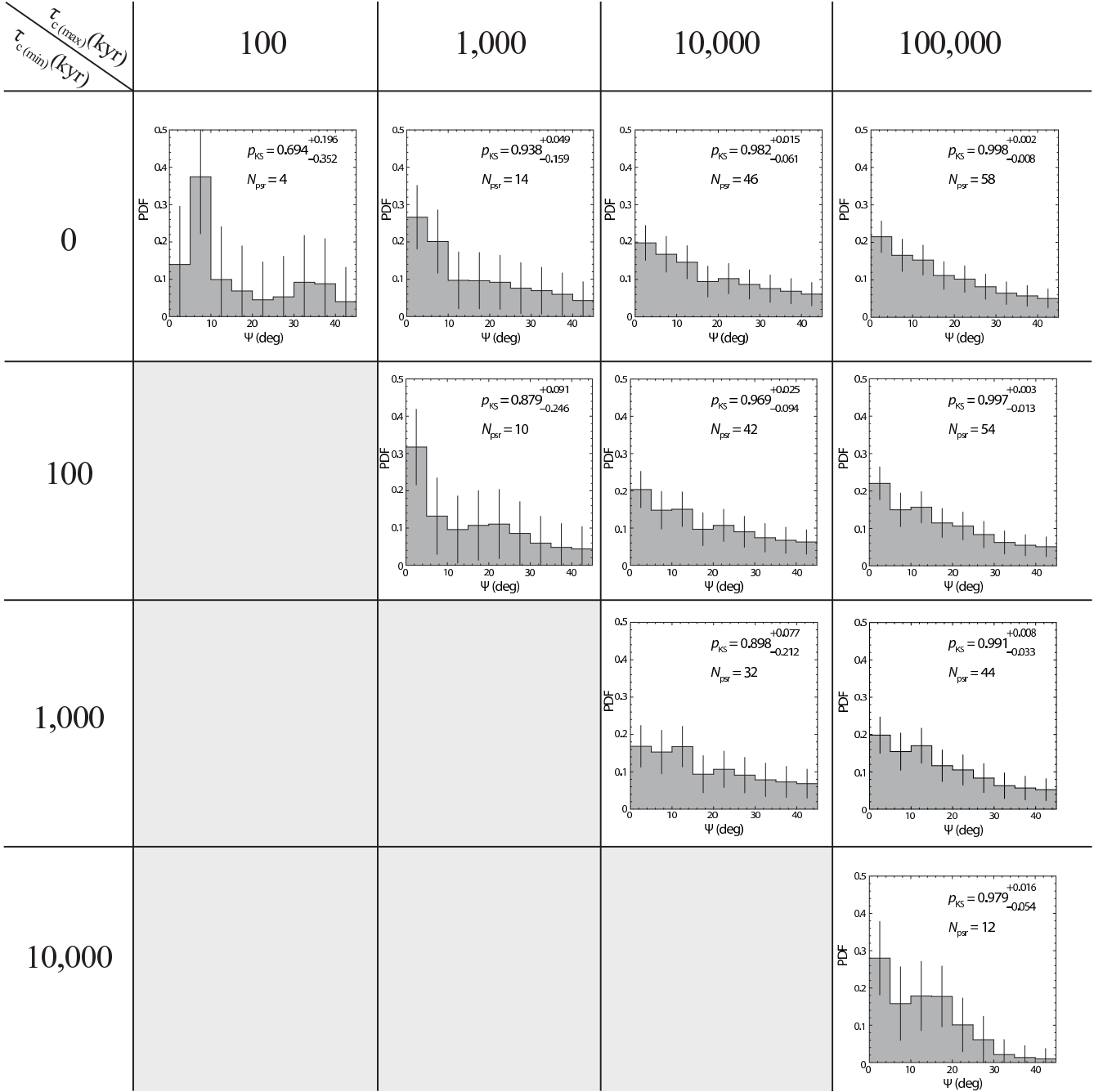} \caption{\label{fig:fig5} Probability distributions of the difference between the position angles of the pulsar spin (${\rm PA}_0$) and velocity (${\rm PA}_{\rm v}$), $\Psi$, for a range of characteristic-age intervals, $\tau_{c({\rm min})}\leq\tau_{\rm c}<\tau_{c({\rm max})}$. Each distribution shown was generated from a large number of MC realisations of ${\rm PA}_0$ and ${\rm PA}_{\rm v}$, based on their published $1\sigma$ errors. The number of spin--velocity angles considered in each distribution is shown as $N_{\rm psr}$, where the total number of pulsars ($N_{\rm psr}=58$) was considered for the distribution covering the entire investigated range of characteristic ages, i.e.~$0\leq \tau_{\rm c}<100,000$ kyr. The error bar shown at each $5^\circ$-bin of $\Psi$ corresponds to the standard deviation of the bin's height across all MC iterations. The probability of rejecting the uniform $\Psi$ distribution, given the observed distribution, calculated with the Kolmogorov--Simirnov (KS) test, is shown as $p_{\rm KS}$. The 1$\sigma$ confidence interval of $p_{\rm KS}$, for each distribution of $\Psi$, was derived from the distribution of the KS statistic across the MC simulation.} 
\end{figure*}

\section{Introduction}
\label{sec:intro}
The alignment between the spin and velocity orientations of pulsars has been motivated both theoretically and observationally. A number of mechanisms have been proposed that predict asymmetric kicks which lead to pulsar birth velocities along the direction of their spin axis (e.g.~Tademaru \& Harisson 1975; Spruit \& Phinney 1998; Cowsik 1998). Moreover, pulsar-population syntheses based on those models predict certain correlations between pulsar observables and the degree of alignment that if present would lend support to the above mechanisms (e.g.~Ng \& Romani 2007; Wang , Lai \& Han 2007; Kuranov, Popov \& Postnov 2009). Observationally, there is strong evidence for spin--velocity alignment in individual pulsars, from X-ray imaging of pulsar-wind nebulae tori (e.g.~the Crab and Vela pulsars; Caraveo \& Mignani 1999; Ng \& Romani 2004; Ng \& Romani 2007), and in samples of pulsars where polarimetric observations can reveal the spin-axis orientation (Johnston et al.~2005; Rankin 2007; Noutsos et al.~2012). As was highlighted in Noutsos et al.~(2012), without available information on the 3D orientation of pulsar velocities and spin-axis directions, statistical studies of pulsar spin--velocity alignment that are based on pulsar proper motions and polarimetric data have to rely on the projected spin-axis and velocity vectors onto the plane of the sky. Apart from the few exceptions mentioned above, this is the case for all pulsars considered in Noutsos et al.~(2012) and in the follow-up study presented here.

Recently, Kuranov, Popov \& Postnov (2009) used Monte Carlo (MC) population synthesis to examine the effect of binary break-up and isolated pulsar kicks on the observed distribution of space-velocity magnitudes and the spin--velocity offset angles. The above authors compared the simulated distributions with those from the observed data sample of Rankin (2007) and Ng \& Romani (2007), which helped them determine valid ranges for physical parameters, like the range of spin--kick-velocity offsets of a pulsar population with various fractions of binary progenitors. However, so far, there has been little attention to the dependence of the observed alignment on pulsar age. Indeed, if there is a mechanism that favours spin--velocity alignment, then one should expect to see such a correlation for young pulsars ($\lesssim 1$ Myr). For older pulsars ($\gtrsim 10$ Myr), the Galactic potential will alter the pulsar proper motions, diluting the effect and causing the correlation to weaken. The typical timescale in which this happens can be estimated as $[GM_{\rm MW}/(\pi R_{\rm MW}^2\delta h)]^{-1/2}\sim 10$ Myr, where $M_{\rm MW}=10^{12}$ M$_\odot$ and $R_{\rm MW}=25$ kpc is the mass and radius of the Milky Way, respectively, and $\delta h=200$ pc is the Galactic plane thickness. Hereafter in this paper, we refer to the above estimate as the ``dynamical time'', $t_{\rm dyn}$. The effect of the galactic potential was perhaps in play in the studies of Johnston et al.~(2005, 2007): the earlier study showed a clear spin--velocity correlation, whereas the follow up work, using a separate, older sample of roughly equal size, showed no significant correlation. The critical difference between the two samples was the pulsar ages and, what is of equal importance, the larger distances of the second sample compared to the first. For the majority of pulsars, their distance is estimated from their measured dispersion measures (DMs) and a model of the free-electron-density distribution. Using the NE2001 electron-density model of \nocite{cl02}Cordes \& Lazio (2002) typically leads to uncertainties of the order of 20\% on the distance (Cordes \& Lazio 2003\nocite{cl03}). Recent investigations by Gaensler et al.~(2008)\nocite{gmcm08} and Schnitzeler (2012)\nocite{sch12} have shown that in a number of cases, e.g.~for high-latitude pulsars, the discrepancy between the distance derived by the model and that obtained through parallax measurements can be even larger. 
Hence, corrections of pulsar proper motions for the Galactic differential rotation are typically less accurate for distant pulsars, which could, again, weaken an intrinsic correlation.

The following sections of this paper investigate the effect of pulsar age on the spin--velocity correlation, using the sample of Noutsos et al.~(2012). Our sample is composed of 58 pulsars for which polarimetric and proper-motion information was available; in a few cases (i.e.~J0534+2200, J1709$-$4429 and J1952+3252) the spin-axis direction was derived from fitting for the orientation of the pulsar wind nebula tori in X-rays. Detailed information about the observations, the data selection and reduction can be found in Noutsos et al.~(2012) and the references therein. The statistical analysis performed on the total sample in our previous paper showed that spin--velocity alignment is favoured at the $99\%$ confidence level. Furthermore, using toy-model simulations, we showed that the distribution of spin--velocity alignment angles (considering polarisation and orthogonal mode ambiguities) resembles more a distribution of truly aligned configurations and less that of those where spin and velocity are orthogonal. Finally, it was shown that the conclusions are not biased by systematics, e.g.~significantly aligned subsets of data or the choice of ${\rm PA}_0$.

In this paper, we examine the dependence of spin-velocity alignment on pulsar age. As a first step, we take the spin-down ages of our pulsar sample ($\tau_{\rm c}=P/(2\dot{P})$) at face value and examine the distributions of the offset angle between spin and velocity, $\Psi$, as a function of $\tau_{\rm c}$. As a following step, we attempt to provide an alternative measure of the pulsar ages by assuming that our pulsars were born close to the Galactic plane (GP): we trace their Galactic trajectories back in time, using their current proper motions and positions and calculate their kinematic ages, $t_{\rm kin}$, as the time interval between their current position and past intersections with the GP. For the sample of pulsars, for which the determination of $t_{\rm kin}$ was possible, we revisit the dependence of spin--velocity alignment on age using the more reliable $t_{\rm kin}$ instead of $\tau_{\rm c}$. Finally, we investigate what effect large uncertainties on pulsar distance have on the determination of kinematic ages and the spin--velocity correlation. Also, the assumptions that pulsars are born with spin periods, $P_0$, that are much shorter than those observed today, and that they spin down by converting their rotational energy to pure magnetic-dipole emission, i.e.~$\dot{P}\propto 1/P$, is also tested for our sample of pulsars.

\begin{figure} 
\vspace*{10pt} 
\includegraphics[width=0.48\textwidth]{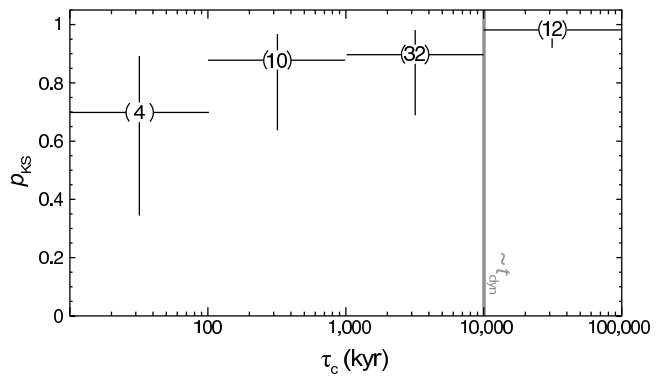} \caption{\label{fig:ksprobs1} Scatter plot of the probability of rejecting the uniform $\Psi$ distribution as a function of characteristic age ($\tau_{\rm c}$), for the $\Psi$ distributions along the main diagonal in Fig.~\ref{fig:fig5}. The data points shown correspond to the values of $p_{\rm KS}$ for the distributions along the main diagonal in Fig.~\ref{fig:fig5}. The number of pulsars considered in the corresponding $\Psi$ distribution is also shown at the position of each marker. The length of time corresponding roughly to the dynamical time of $\sim 10$ Myr is shown with a thick, vertical grey line (see Section~\ref{sec:intro}).} 
\end{figure}

\section{Data Analysis}
\label{sec:datanal} 
The statistical procedure followed in this paper is similar to the one of Noutsos et al.~(2012). Firstly, based on the measurement uncertainties on the position angle of the sky-projected pulsar velocity, ${\rm PA}_v$, and that of the sky-projected spin axis, ${\rm PA}_0$, we generated a large number of MC data sets, each having equal size to the original sample of 58 pulsars. The MC data sets were generated by randomly drawing values from 58 Gaussian distributions with means and standard deviations equal to the published PA values and their uncertainties, respectively. Secondly, for each of the data sets we calculated the values of $\Psi={\rm PA}_0-{\rm PA}_{\rm v}$, while considering the polarisation ambiguity (${\rm PA}_0\pm 180^\circ$) and the ambiguity due to orthogonal-mode emission (${\rm PA}_0\pm 90^\circ$). Due to these ambiguities, the minimum difference between ${\rm PA}_0$ and ${\rm PA}_{\rm v}$ can be defined inside the interval, $-45^\circ\leq\Psi\leq 45^\circ$; but since we are only interested in the absolute offset between the spin and velocity vectors, we only considered the absolute value of the above difference. Hence, all values of $\Psi$ were distributed inside the interval, $0^\circ\leq\Psi\leq 45^\circ$. During the last step, we quantified the degree of spin--velocity correlation, given the observed $\Psi$ distributions, using the one-sample Kolmogorov--Smirnov (KS) test. This test provides, via its test statistic, $D$, the probability that our observed data set of $\Psi$ values is drawn from an arbitrary theoretical distribution, e.g.~the uniform distribution, $p(\Psi)=1/45$ deg$^{-1}$. Throughout this paper, we quote the probability of rejecting the uniform distribution under the KS test, $p_{\rm KS}$: this is defined as 1 minus the probability that the observed $\Psi$ distribution is drawn from a uniform one. In the final step, we generated the cumulative histograms of all MC data sets by summing the values in each bin and calculated the value of $p_{\rm KS}$ corresponding to the mean value of $D$ from all MC runs.

We examined the distributions of $\Psi$ and the corresponding $p_{\rm KS}$ values for different subsets of our pulsar sample, which were defined according to intervals in $\tau_{\rm c}$. The considered range of characteristic ages was $\tau_{\rm c}\leq 100$ Myr, which contained the entire pulsar sample. The range was split into intervals with sizes ranging from one to four orders of magnitude in $\tau_{\rm c}$. Fig.~\ref{fig:fig5} shows the tabulated distributions of $\Psi$, each distribution corresponding to that subset of pulsars which have $\tau_{\rm c}$ in the interval whose lower and upper bounds are shown in kyr along the first column and top row, respectively. The number of pulsars belonging to each age interval is also quoted with each distribution. In addition, to visualise a possible trend in $p_{\rm KS}$ with increasing $\tau_{\rm c}$, we show in Fig.~\ref{fig:ksprobs1} the $p_{\rm KS}$ values of the distributions along the main diagonal of Fig.~\ref{fig:fig5}. 

At first glance, all $\Psi$ distributions appear non-uniform with most of the $\Psi$ values distributed below 20$^\circ$. In general, the incremental addition of older pulsars is accompanied by increasing $p_{\rm KS}$ values, i.e.~$p_{\rm KS}(\tau_{\rm c}<100 \ {\rm kyr})<p_{\rm KS}(\tau_{\rm c}<1,000 \ {\rm kyr})<p_{\rm KS}(\tau_{\rm c}<10,000 \ {\rm kyr})$, etc. The distribution that was generated from all pulsars in our sample shows the highest probability of rejecting uniformity, with the KS test reporting $p_{\rm KS}\approx 99.8$\%. However, the probabilities corresponding to discrete intervals of $\tau_{\rm c}$, as is shown in Fig.~\ref{fig:ksprobs1}, show no correlation between $p_{\rm KS}$ and $\tau_{\rm c}$. Interestingly, we also have $p_{\rm KS}(\tau_{\rm c}\geq 0 \ {\rm kyr})>p_{\rm KS}(\tau_{\rm c}\geq 100 \ {\rm kyr})>p_{\rm KS}(\tau_{\rm c}\geq 1,000 \ {\rm kyr})$, etc. Overall, there is no evident trend towards more uniform distributions of $\Psi$ (i.e.~smaller values of $p_{\rm KS}$) with increasing characteristic age. Nevertheless, it is true that the histograms for the very young pulsars ($<100$ kyr), those with ages between 100 and 1,000 kyr, as well as the very old pulsars ($\geq 10,000$ kyr) are derived from a small sample of $\sim 10$ pulsars. Therefore, our confidence in the reported probabilities for those cases is generally lower than for the rest of the distributions, which is also reflected in the size of the errors on $p_{\rm KS}$: these were derived from the 68\% confidence limits (CLs) of the KS statistic, $D$. The exception to this is the distribution for $1,000\leq \tau_{\rm c}<10,000$ kyr, which is based on a sizeable sample of 32 pulsars but has also a large lower uncertainty on $p_{\rm KS}$ (i.e.~23\% lower error). 

In conclusion, if we take the above results at face value, we can exclude (at the $90\%$ confidence level) that spin and velocity orientations are completely uncorrelated, for the entire investigated range of characteristic ages. Yet, if there is an age-dependent smearing of the spin--velocity alignment, it may only become effective beyond $\sim 10^8$ y, assuming $\tau_c$ to be reliable (but read Section~\ref{sec:pulkinage}). However, as was noted in the introduction, this is hard to reconcile with the dynamical time of $\sim 10$ Myr, being an order of magnitude shorter that this limit.

\section{Pulsar Kinematic Ages} 
\label{sec:pulkinage} 
Pulsar age is a critical parameter in our investigation of whether pulsar spin vectors tend to be aligned with their velocities for young pulsars. However, except for the few cases where historical pulsar--supernovae associations can provide good estimates of pulsar ages, the only other measure we possess of a pulsar's age is its spin-down or characteristic age ($\tau_{\rm c}\propto P/\dot{P}$). Unfortunately, in many cases it can be readily seen that the characteristic age is an unreliable estimate of the true age, $t_{\rm true}$: e.g.~tracing the outward proper motions of PSR B1951+32 and PSR B0538+2817 from the centres of their respective supernova remnants (SNRs) has revealed that their true ages are $t_{\rm true}\sim 0.5\tau_{\rm c}$ and $\sim 0.05\tau_{\rm c}$, respectively (Migliazzo et al.~2002; Kramer et al.~2003). This is largely due to the implicit assumptions of the above expression for $\tau_{\rm c}$, which are that pulsars are born with periods much shorter than those observed today and that they spin down by converting their rotational energy to low-frequency magnetic-dipole radiation, while maintaining a constant magnetic-field. An additional source of systematic uncertainty in the determination of $\tau_{\rm c}$ comes from the ``Shklovskii effect'', whereby the apparent spin-down rate, $\dot{P}$, is contaminated with kinematic contributions like the relative transverse velocity between the pulsar and the solar system barycentre (Camilo et al.~1994\nocite{ctk94}); the Galactic differential acceleration and vertical acceleration introduce additional $\dot{P}$ components (Damour \& Taylor 1991\nocite{dt91}). The Shklovskii effect on the measured $\dot{P}$, and hence on $\tau_{\rm c}$, becomes important for high-velocity, nearby pulsars, whereas Galactic differential acceleration and vertical acceleration are important for distant and high latitude pulsars, respectively. In general, the contribution of these dynamical effects to the value of $\dot{P}$ is negligible for slow-spinning, non-recycled pulsars. However, for nearby, high-velocity millisecond pulsars, with values of $\dot{P}$ that are $\sim 5$ orders of magnitude smaller, the above effects may account for a significant fraction of the measured $\dot{P}$ (see e.g.~Section 8.2.4 of Lorimer \& Kramer 2005\nocite{lk05}).

Given the aforementioned uncertainties associated with the determination of pulsar ages based on spin down, it is important to explore alternative methods of determining those ages that are independent of a spin-down model. To date, there are only a few non-recycled pulsars whose ages have been determined through independent methods. Some of them exhibit a large discrepancy, up to an order of magnitude, between $\tau_{\rm c}$ and the independent estimates. The discrepancy can reveal $\tau_{\rm c}$ as being either an over- or under-estimate of the true age of a pulsar. For instance, the true age of PSR J0538+2817 is roughly an order of magnitude younger that its characteristic age (Kramer et al.~2003\nocite{klh+03}). Conversely, the westward proper motion of PSR J1801$-$2451 suggests that the pulsar's true age must be significantly higher than its characteristic age, if the pulsar is associated with in the SNR G5.4$-$1.2 (Gaensler \& Frail 2000\nocite{gf00}; also see Zeiger et al.~2008\nocite{zbcg08}). Furthermore, the outwards proper motion of PSR J1932+3252 from the geometric centre of its parent SNR CTB 80 implies a true age of $\sim 51$ kyr for this pulsar (Zeiger et al.~2008\nocite{zbcg08}). However, these examples are of typically young pulsars with known supernova associations; it is quite possible that a large number of older pulsars, for which we have no independent clues about their ages, have significantly different true ages to their characteristic ones.

Under the assumption that pulsars spin down entirely due to electromagnetic radiation, it is generally assumed that the spin-down rate is proportional to some power law of the spin rate: this is expressed as
\begin{equation} 
\label{eq:spindown} 
\dot{\nu}=-K\nu^n, 
\end{equation}
where $\nu=1/P$ is the spin frequency; $K$ is a proportionality factor, often assumed to be constant, that depends on the physical properties of the pulsar, i.e.~the moment of inertia, the magnetic-dipole moment and the inclination angle between the spin- and magnetic axes; $n$ is the pulsar braking index (Manchester \& Taylor 1977\nocite{mt77}; Shapiro \& Teukolsky 1983\nocite{st83}). The general solution of the above differential equation for the true age of a pulsar, $t_{\rm true}$, is
\begin{equation} 
\label{eq:truage} 
t_{\rm true}=\frac{P}{(n-1)\dot{P}}\left[1-\left(\frac{P_0}{P}\right)^{n-1}\right], \ \ {\rm if} \ n \neq 1 \end{equation}
Here, we have implicitly assumed that apart from $K$, the braking index also remains constant with time. However, the observed trends in a number of pulsar properties with age, e.g.~the magnetic inclination angles, suggest that $K$ and $n$ are functions of time, implying that pulsar magnetic fields decay with time and/or that their magnetic axes tend towards alignment with their spins, as they age (Tauris \& Manchester 1998\nocite{tm98}; Young et al.~2010\nocite{ycbb10}). If this is true, then the above solution is invalid; in that case, Eq.~\ref{eq:spindown} is generalised as
\begin{equation} 
\dot{\nu}(t)=-K(t)[\nu(t)]^{n(t)}, 
\end{equation}
which if solved would give the true age of a pulsar but whose form is unknown and likely varies between pulsars. According to Vranesevic \& Melrose~(2011)\nocite{vm11}, if the evolutionary motion of pulsars on the $P$--$\dot{P}$ diagram constitutes a current that is a conserved quantity, i.e.~the observed distribution is maintained in time, then pulsar spin-down can be generally expressed as $\ddot{P}=-dC(P)/dP$, where $C(P)$ is the unknown potential responsible for the pulsar current. In the specific case of pulsar evolution under a constant $n$, the above generalised spin-down law reduces to Eq.~\ref{eq:spindown}. 

Under the usual assumptions of $P_0/P=0$ and $n=3$, Eq.~\ref{eq:truage} simply reduces to the {\em characteristic age} or {\em spin-down age} of a pulsar
\begin{equation} 
\label{eq:charage} 
\tau_{\rm c}=\frac{P}{2\dot{P}} \ . 
\end{equation}
However, all reliably measured braking indices to date are less than 3; amongst them there are extreme cases where $n\approx 1$, like that of the Vela pulsar and PSR J1734$-$3333 (Espinoza et al.~2011\nocite{elk+11}). Hence, if these pulsars were born fast and have evolved under a constant braking index, then $\tau_{\rm c}$ would be an underestimate of their true age. The age of pulsars with $n=2$ would then be closer to   
\begin{equation} 
\label{eq:charage2} 
\tau_2=2\tau_{\rm c} 
\end{equation}
where we have assumed again that $P_0/P=0$ in Eq.~\ref{eq:truage}. Furthermore, for pulsars that have evolved with $n=1$ even $\tau_2$ would be an underestimate. For this special case, Eq.~\ref{eq:spindown} has a special solution, i.e.
\begin{equation} 
\label{eq:charage1a} 
\ \ \ \ \ \ \ \ \ \ \ \ \ \ \ \ t_{\rm true}=2\tau_{\rm c}\ln\left(\frac{P}{P_0}\right), \ \ {\rm if} \ n=1
\end{equation}
which cannot be simplified further. Nevertheless, we can arbitrarily choose $P_0$ to be very small as to have again $P_0/P\approx 0$ for non-recycled pulsars. Choosing $P_0=1$ ms, close to the break-up limit for a neutron star (e.g.~Chakrabarty et al.~2003\nocite{cmm+03}), we then have
\begin{equation} 
\label{eq:charage1} 
\tau_1=2\tau_{\rm c}\ln\left(\frac{P}{1 \ {\rm ms}}\right) 
\end{equation}
where $P$ is in ms. 

Finally, it is easy to show that pulsar age, as defined by Eq.~\ref{eq:truage}, increases monotonically with decreasing $n$, for any value of $0<P_0<P$ and any $n$ (this also remains true around the special case of $n=1$). Consequently, for every pulsar, $\tau_1>\tau_2>\tau_{\rm c}$: i.e.~larger values of $n$ mean that the pulsar spins down more rapidly.

\subsection{Characteristic Age Bias due to $P_0$ and $n$}
\label{subsec:p0nbias}
Invoking Eqs.~\ref{eq:spindown},\ref{eq:truage} and \ref{eq:charage}, the ratio between $t_{\rm true}$ and $\tau_{\rm c}$, at an arbitrary time, $t$, can be expressed as a function of the ratio, $r_1$, measured at the present time, $t_1$:
\begin{equation} 
r(t)=\frac{t_{\rm true}}{\tau_{\rm c}}=r_1\left(\frac{1+\chi/t_1}{1+\chi/t}\right), 
\end{equation}
where $\chi=P_0^{n-1}/[K(n-1)]$. Since $P_0$ is usually unknown, we can express the constant $\chi$ as a function of $r_1$ and $t_1$ by solving Eq.~\ref{eq:spindown} and replacing the expression for $P(t)$ in $\tau_{\rm c}=P/(2\dot{P})$:
\begin{equation} 
\label{eq:ageratio} 
r(t)=\frac{\rho}{1+(\rho/r_1-1)\dfrac{t_1}{t}}, 
\end{equation}
where we have set $\rho=2/(n-1)$.

As an example, we can use Eq.~\ref{eq:ageratio} to predict the expected discrepancy between $t_{\rm true}$ and $\tau_{\rm c}$, for a particular pulsar, in future. As was mentioned earlier, PSR J0538+2817 has $\tau_{\rm c}=600$ kyr and $t_1=30$ kyr, which translates to $r_1=0.05$. Assuming $n=3$, according to Eq.~\ref{eq:ageratio} the ratio $t_{\rm true}/\tau_{\rm c}$ becomes $r\approx 0.95$, when $t_{\rm true}=10$ Myr. This result is expected, because, even though today's period of PSR J0538+2817 is comparable to its birth period, as the pulsar becomes older the assumption that $P_0\ll P$ becomes more valid. Moreover, if $n=2$, Eq.~\ref{eq:ageratio} gives $r=1.8$, for $t=10$ Myr. In the more extreme case of the Vela pulsar, the braking index could be as low as 1.2 ($1\sigma$ lower bound), which would mean $r=6.3$ at $t=10$ Myr. Finally, for the special case where $n=1$, we simply have $r(t)=r_1(t/t_1)$; for PSR J0538+2817, it would mean that $r=16.7$. So, the age bias in $\tau_{\rm c}$ due to the unknown birth period of pulsars is less important for older pulsars but can still be as big as an order of magnitude for young pulsars (e.g.~PSR J0538+2817). On the other hand, the bias due to the braking index is far more important for old pulsars. If several of the old pulsars in our sample ($>10,000$ kyr) have evolved under a constant braking index that has been close to 1, like that of the Vela pulsar at present, then their characteristic ages could be as much as 10 times younger than their true ages (see Section~\ref{subsec:brakindx}).

In the worst case, both of the above effects could be in operation for the pulsars in our sample, which would mean that if we replaced $\tau_{\rm c}$ with $t_{\rm true}$, a number of pulsars would belong to different age intervals in Fig.~\ref{fig:fig5}. More specifically, middle-aged pulsars according to $\tau_{\rm c}$(=$100-10,000$ kyr) would be shifted to the distributions for $t_{\rm true}<100$ kyr, due to their large $P_0$. Similarly, some of those middle-aged pulsars would belong to the distribution for $t_{\rm true}>10,000$ kyr, due to $n<3$. In the following paragraphs we consider an alternative method of estimating the age of the pulsars in our sample and re-examine the spin--velocity alignment as a function of age, based on those new estimates.

\subsection{Methodology} 
\label{subsec:method}
It is generally believed that short-lived, massive OB stars, that dwell near the GP, are the likely progenitors of pulsars (e.g.~\nocite{agk86}Amnuel, Guseinov \& Kustamov 1986; but see also Blauw 1985\nocite{bla85}). It is therefore a reasonable assumption that pulsars are born at Galactic heights of $|z_{\rm birth}|\ll R_{\rm MW}$. Based on this assumption, pulsar ages can be estimated using a kinematic analysis, by calculating the length of time required for a pulsar to travel between its location at birth and that at present, through the gravitational potential of the Galaxy. Such a method has been discussed in the literature by various investigators (e.g.~Lyne, Anderson \&  Salter 1982\nocite{las82}; Brisken et al.~2003\nocite{bfg+03a}). In general, pulsar ages derived from such kinematic analyses are free of the biases discussed in the previous section and, thus, can be more reliable than characteristic ages. In the following paragraphs, we describe the method used in this paper to derive ``kinematic ages'', hereafter $t_{\rm kin}$, for the pulsars in our sample. Our goal is to re-examine the dependence of spin--velocity alignment on pulsar age, this time using the more reliable kinematic ages.  

\subsubsection{The pulsar trajectories through the Galaxy}
The calculation of a pulsar's path through the Galaxy, e.g.~between its birth site and its present position, requires a numerical integration of the equations of motion through a Galactic gravitational potential. Primarily, this calculation requires knowledge of the following pulsar observables, which form the initial conditions for the integration, at present time:
\begin{itemize}
\item[--] The pulsar's position in Galactic coordinates, $(l,b)$.
\item[--] The pulsar's distance, $d$. 
\item[--] The pulsar's proper motion, $(\mu_l,\mu_b)$.
\end{itemize}
Proper motions have been measured for hundreds of pulsars and can be found in the literature (e.g.~Hobbs et al.~2005\nocite{hllk05}). On the other hand, pulsar distances are only accurately known for a few tens of pulsars, from parallax measurements and SNR associations. Nevertheless, the distances of most pulsars can be crudely estimated via their dispersion measures (DM) and a model of the Galactic free-electron density distribution (Cordes \& Lazio~2002\nocite{cl02}; Schnitzeler~2012\nocite{sch12}).

In addition, the numerical integration requires an assumption about the Galactic gravitational potential. Several potentials exist in the literature (e.g.~Kenyon et al.~2008\nocite{kbg+08}; Kuijken \& Gilmore 1989\nocite{kg89}). In this work, we have chosen the potential of Paczynski (1990)\nocite{pac90}. These potentials are generally expressed in right-handed Galactocentric Cartesian coordinates (GCC), $(x,y,z)$, where the $x$ axis is directed from the Sun to the Galactic centre (GC). Since the pulsar parameters are always measured with respect to the solar system barycentre (SSB), it is convenient to express them also in GCC. The pulsar's present velocity in GCC, as measured with respect to the solar system barycentre (SSB), is then given by
\begin{align} 
\label{eq:velocity} 
\nonumber v_x= & v_r \sin l  \cos b + \mu_l d \cos l   - \mu_b d \sin l \sin b  \\ v_y= & -v_r \cos l  \cos b + \mu_l d \sin l   + \mu_b d \cos l \sin b  \\ \nonumber v_z= & v_r \sin b + \mu_b d\cos b 
\end{align}
The initial conditions for the numerical integration critically depend on the unknown radial velocity, i.e.~$\boldsymbol{v}_{\rm SSB}(v_{\rm r})$. Like with $z_{\rm birth}$, we have also numerically explored a range of possible $v_r$ values. The choice of this range was motivated by the work of Hobbs et al.~(2005)\nocite{hllk05}, who compiled a large catalogue of 233 pulsar proper motions, derived from timing-parallax measurements. Their work showed that the distribution of observed magnitudes of pulsar transverse velocities, along one dimension (1D distribution), is roughly contained within 1,000 km s$^{-1}$. Hence, in this work we have explored the range $|v_r|\leq 500$ km s$^{-1}$, where positive $v_r$ is defined as being directed away from the observer. As before, we defined a fine grid with $\Delta v_r = 1$ km s$^{-1}$ and calculated the pulsar's trajectory for each of the grid values.

\begin{figure*} 
\vspace*{10pt} 
\includegraphics[width=1.0\textwidth]{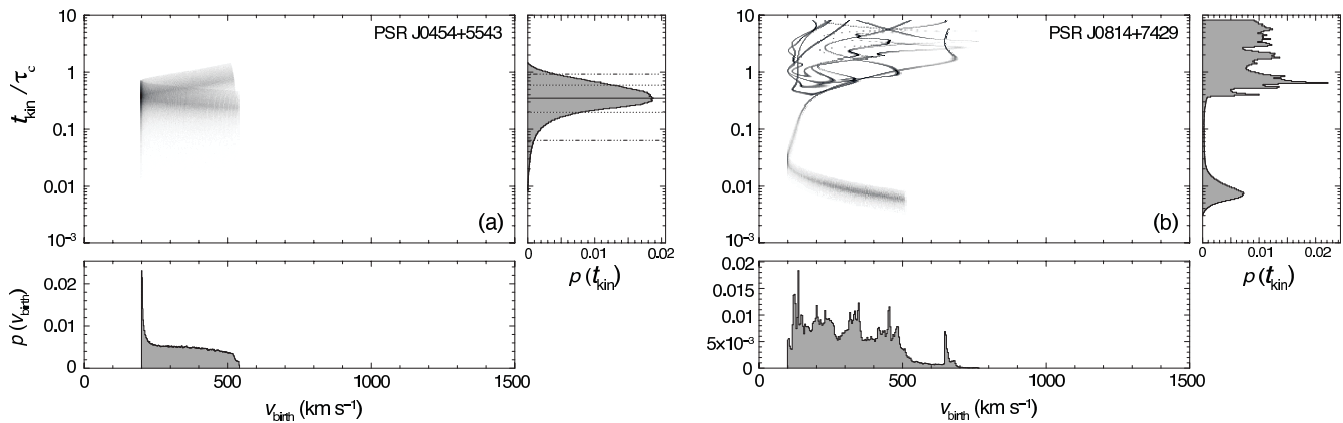} 
\caption{\label{fig:pdfs} Posterior PDFs derived from our MC analysis for (a) PSR J0454+5543 and (b) PSR J0814+7429. (Central panels) The posterior probability, $p(v_{\rm birth},t_{\rm kin})$, from a large number of MC realisations of $z$ and $v_{\rm birth}$, randomly drawn from the exponential distribution of Reed (2000) and the Maxwellian distribution of Hobbs et al.~(2005), respectively. (Bottom panels) The marginalised posterior probability, $p(v_{\rm birth})$, derived from $p(v_{\rm birth},t_{\rm kin})$ by integration over all $t_{\rm kin}$. (Right panels) The marginalised posterior probability, $p(t_{\rm kin})$, normalised by $\tau_{\rm c}$, derived from $p(v_{\rm birth},t_{\rm kin})$ by integration over all $v_{\rm birth}$. For PSR J0454+554, the simple form of $p(t_{\rm kin})$ allows us to quote the 68\% (dotted line) and 95\% (dashed-dotted line) confidence intervals around the peak value (solid line). For PSR J0814+7429, the $p(t_{\rm kin})$ has a complex shape with multiple peaks of comparable probability. Hence, we cannot confidently quote a value for $t_{\rm kin}$; in this case, only the age at PDF maximum is reported, also stating that this is an unreliable age estimate (see Table~\ref{tab:tab2}).} 
\end{figure*}

We can now numerically perform the integration of the equations of motion from the present values of $\boldsymbol{r}_{\rm GCC}(x,y,z)$ and $\boldsymbol{v}_{\rm SSB}(x,y,z)$ to the values at birth, $\boldsymbol{r}_{\rm GCC}(x_{\rm birth},y_{\rm birth},z_{\rm birth})$ and $\boldsymbol{v}_{\rm SSB}(x_{\rm birth},y_{\rm birth},z_{\rm birth})$. 

As was mentioned earlier, we assume that pulsars are born very close to the Galactic plane. In this work, we have numerically explored a range of birth heights, with $|z_{\rm birth}|\leq 100$ pc. The chosen range was motivated by the work of Reed (2000)\nocite{ree00}, who used the observed latitudes and magnitudes of OB stars to determine the scale height of their distribution to $h_{\rm OB}\sim 45$ pc. In his work, Reed assumed that the optical extinction of the Galactic disc in the $B$-band, within 100 pc, is 2 magnitudes kpc$^{-1}$. In order to numerically simulate a continuous sampling of the chosen range in $z_{\rm birth}$, we formed a very fine grid of birth heights with $\Delta z_{\rm birth}=0.2$ pc. Each intersection of the pulsar's trajectory with grid elements was considered a possible birth site. The birth location along the other two coordinates, i.e.~$(x_{\rm birth},y_{\rm birth})$, was left unconstrained.

Finally, a number pulsars, e.g.~PSR J1900$-$2600 ($\tau_{\rm c}\sim 100$ Myr), may have crossed the Galactic plane several times in their lifetime before reaching their current position. If unrestricted, and depending on the value of $v_r$, our simulation would allow pulsars that are bound to the Galactic gravitational potential to cross the GP indefinitely. Therefore, we decided to place a strict upper limit on the length of our integration. In particular, for each pulsar, we considered all possible intersections with the GP within $\tau_1$. Hence, our simulation places an artificial upper limit on the kinematic age of each pulsar, i.e.~$t_{\rm kin}\leq \tau_1$. This limit corresponds to the case in which the pulsars are born with $P_0=1$ ms and spin down exponentially on a characteristic timescale of $\tau_2=2\tau_{\rm c}$ (see Eq.~\ref{eq:charage1}). Given the spin periods of the 58 pulsars in our sample, this limit ranges between $\tau_1\sim 7\tau_{\rm c}$ and $15\tau_{\rm c}$, with 52 out of the 58 pulsars having $\tau_1>10\tau_{\rm c}$. Therefore, we deemed $\tau_1$ a long-enough time scale for the purpose of our simulation.   
In Section~\ref{sec:multisect}, we discuss in more detail the implications of having multiple intersections with the GP. 

Following the formation of the above grids, for each pulsar we calculated $t_{\rm kin}$ for all $(v_r,z_{\rm birth})$ pairs that produced one or more intersections within $\tau_1$. Up to this point, our method does not consider the actual probability distributions of $v_r$ and $z_{\rm birth}$. In reality, those distributions are far from uniform, as it is more likely that a pulsar is born closer to the plane than away from it; similarly, it is more likely that a pulsar is travelling at $\sim 100$ km s$^{-1}$ than at $1,000$ km s$^{-1}$. At the next step of our simulation, we consider those two probability distributions.

\subsubsection{The distribution of birth heights}
The work by Reed (2000)\nocite{ree00} has shown that the probability density of the birth heights of OB stars, the likely progenitors of pulsars, falls exponentially with $z$: i.e.
\begin{equation} 
\label{eq:density} 
\mathfrak{R}(z)\propto \exp(-|z|/h_{\rm OB}) 
\end{equation}
where the exponential scale height, $h_{\rm OB}$ was determined in that work to be 45 pc. In order to incorporate this probability distribution into our simulation, we first generated a large number of values of $z$ from $\mathfrak{R}(z)$, allowing only $|z|<100$ pc. It is easy to show that the considered range of birth heights covers $\approx 90$\% of $\mathfrak{R}(z)$. For each generated $z$ value, we found its nearest neighbour in the grid of $z_{\rm birth}$ and kept all the corresponding values of $t_{\rm kin}$ that had been previously calculated for that $z_{\rm birth}$. After doing this, we are left with a large number of $(v_r,z_{\rm birth})$ pairs and corresponding $t_{\rm kin}$ values, appropriately weighted based on Eq.~\ref{eq:density}. 

\subsubsection{The distribution of birth velocities}
In the previous paragraph, the consideration of the distribution of birth heights ensured that smaller values of $|z_{\rm birth}|$ are weighted higher than larger ones. However, up to this point of our simulation all pulsar velocities, $|v_r|\leq 500$ km s$^{-1}$, are treated as equally probable. The distribution of pulsar radial velocities, $v_r$, is unknown, so it is not possible to directly follow the previous procedure, as was done for $z_{\rm birth}$. Nevertheless, based on the proper motions of 233 pulsars, Hobbs et al.~(2005) conclude that the distribution of pulsar velocities at birth, $v_{\rm birth}$, can be fitted with a single Maxwellian:
\begin{equation} 
\label{eq:HobbsDist} 
\mathfrak{H}(v)=\sqrt{\frac{2}{\pi}}\frac{v^2}{\sigma^3} \exp(-\frac{v^2}{2\sigma^2}) 
\end{equation}
where $\sigma=265$ km s$^{-1}$ is the standard deviation of the observed 1D pulsar velocities, in either the longitudinal or latitudinal direction. Before we can use $\mathfrak{H}(v_{\rm birth})$ with our data, we need to transform the sample of $v_r$ values, from the previous step, to $v_{\rm birth}$. This transformation entails the following two steps:
\begin{itemize}
\item[--] Correction of $\boldsymbol{v}_{\rm SSB}$ for the motion of the local standard of rest (LSR), which affects the presently observed pulsar velocities. Hence, for each value of $v_{\rm r}$, we subtracted the velocity of the LSR, $\boldsymbol{V}_\odot$, which we assumed to be azimuthal at the Solar circle (Galactocentric distance of $R_{\odot}=8.5$ kpc), directed clockwise, seen from the North: i.e.
\begin{equation} \boldsymbol{v}_{\rm LSR}(x,y,z) = \boldsymbol{v}_{\rm SSB}(x,y,z) + V_{\odot}\boldsymbol{\hat{y}} 
\end{equation} where we have chosen $V_{\odot}=-225$ km s$^{-1}$, following the value used in Harrison et al.~(1993)\nocite{hla93}.
\item[--] Correction of the pulsar's velocity at the birth location, $(x_{\rm birth},y_{\rm birth},z_{\rm birth})$, for the in-situ Galactic rotation, $\boldsymbol{V}_{\rm MW}(x_{\rm birth},y_{\rm birth})$. This calculation requires first the determination of the pulsar's birth location, which was performed by numerical integration of the equations of motion using $\boldsymbol{v}_{\rm LSR}$. The resulting velocity is the pulsar's velocity at birth, relative to the local ISM: 
\begin{equation} 
\boldsymbol{v}_{\rm birth} = \boldsymbol{v}_{\rm LSR}(x_{\rm birth},y_{\rm birth}) + \boldsymbol{V}_{\rm MW}(x_{\rm birth},y_{\rm birth}) 
\end{equation}
whose magnitude, $v_{\rm birth}$, is always positive. It should be noted that in the above calculations we have assumed a flat rotation curve for the Galaxy, i.e.~no radial or vertical dependence of $V_{\rm MW}$. 
\end{itemize}

\subsubsection{The posterior probability distributions}
The above procedure transforms the original $(v_r,z_{\rm birth})$ pairs to pairs of $(v_{\rm birth},z_{\rm birth})$. In the last step, we first divided the ranges of $v_{\rm birth}\in [v_{\rm birth}^{\rm min},v_{\rm birth}^{\rm max}]$ and $t_{\rm kin}\in [t_{\rm kin}^{\rm min},t_{\rm kin}^{\rm max}]$, for each pulsar, into 1,000 bins: on the $v_{\rm birth}$--$t_{\rm kin}$ plane, these form a 2D grid with bin size, $\delta v_{\rm birth} \times \delta t_{\rm kin}$, where $\delta v_{\rm birth}=\tfrac{1}{1,000}(v_{\rm birth}^{\rm max}-v_{\rm birth}^{\rm min})$ and $\delta t_{\rm kin}=\tfrac{1}{1,000}(t_{\rm kin}^{\rm max}-t_{\rm kin}^{\rm min})$. Next, we generated a large number of values of $v$ from $\mathfrak{H}(v)$. Note that the range of $v_{\rm birth}$, although implicitly restricted by the chosen range of $v_r$, can vary significantly between pulsars with different locations and proper motions. For each of the randomly drawn values of $v$, we calculated the number density of the data from our simulation that fell into each of the 2D rectangular bins. The resulting 2D density plot constitutes the 2-dimensional (2D), posterior probability density function (PDF): 
\begin{align}
\label{eq:2Dpdf} 
\nonumber p(v_{\rm birth},t_{\rm kin})=H(&\tau_1 - t_{\rm kin})\int_{-100 \ {\rm pc}}^{{+100 \ {\rm pc}}}dz_{\rm birth}\times \\
&\times \mathfrak{L}(v_{\rm birth},t_{\rm kin})\mathfrak{H}(v_{\rm birth})\mathfrak{R}(z_{\rm birth}), 
\end{align} 
where $\mathfrak{L}(v_{\rm birth},t_{\rm kin})$ is the likelihood function that depends implicitly on the model parameters, $v_r$ and $z_{\rm birth}$ --- becoming zero outside their investigated ranges --- and $H(x)$ is the Heaviside function that simply indicates that all solutions corresponding to $t_{\rm kin}>\tau_1$ are rejected. Fig.~\ref{fig:pdfs} shows two examples of grayscale density maps of $p(v_{\rm birth},t_{\rm kin})$, where $t_{\rm kin}$ is normalised by $\tau_{\rm c}$. Notice that we derived $p(v_{\rm birth},t_{\rm kin})$ by marginalising $z_{\rm birth}$ over the considered range.

A direct product of the above simulation were the marginalised posterior PDFs for $v_{\rm birth}$ and $t_{\rm kin}$, 
\begin{align}
\label{eq:marginalPDFs} 
\nonumber 
& p(v_{\rm birth})=\int_0^\infty p(v_{\rm birth},t_{\rm kin})dt_{\rm kin} \\ 
& p(t_{\rm kin})=\int_0^\infty p(v_{\rm birth},t_{\rm kin})dv_{\rm birth} 
\end{align} 
which are shown in the bottom and right-hand-side panels of Fig.~\ref{fig:pdfs}, respectively.

Evidently, despite the well-defined form of $\mathfrak{H}(v_{\rm birth})$ (i.e.~the single-mode Maxwellian of Hobbs et al.), the shape of $p(v_{\rm birth})$ differs significantly from the prior distribution. The main reason for this is the multiplicity of solutions for $t_{\rm kin}$, for a given $v_{\rm birth}$, which arises from the parametric pair of equations that relates $v_{\rm birth}$ and $t_{\rm kin}$: i.e. 
\begin{align}
\label{eq:parametric} 
\nonumber v_{\rm birth}&=v_{\rm birth}(z_{\rm birth}, v_r) \\ 
t_{\rm kin}&=t_{\rm kin}(z_{\rm birth}, v_r) 
\end{align} 
And, of course, the same is true for $p(t_{\rm kin})$, which means that we cannot recover the intrinsic PDF of $t_{\rm kin}$. It is important to clarify that the multiple solutions to which we are referring, here, do not arise from multiple intersections with the GP but are the result of the multiple possibilities for $(v_{\rm birth},z_{\rm birth})$ that lead to the same $t_{\rm kin}$, and, vice versa, the multiple $(t_{\rm kin},z_{\rm birth})$ that lead to the same $v_{\rm birth}$. Allowing for multiple intersections adds an additional degree of complexity to the derived solutions, which is discussed in Section~\ref{sec:multisect}. 

Finally, it should be stressed that, given the lack of prior information, e.g.~association of the birth location with a spiral arm or high-density region, all valid solutions of Eq.~\ref{eq:parametric} in our simulation were equally weighted.

\begin{figure*} 
\vspace*{10pt} 
\includegraphics[width=1.0\textwidth]{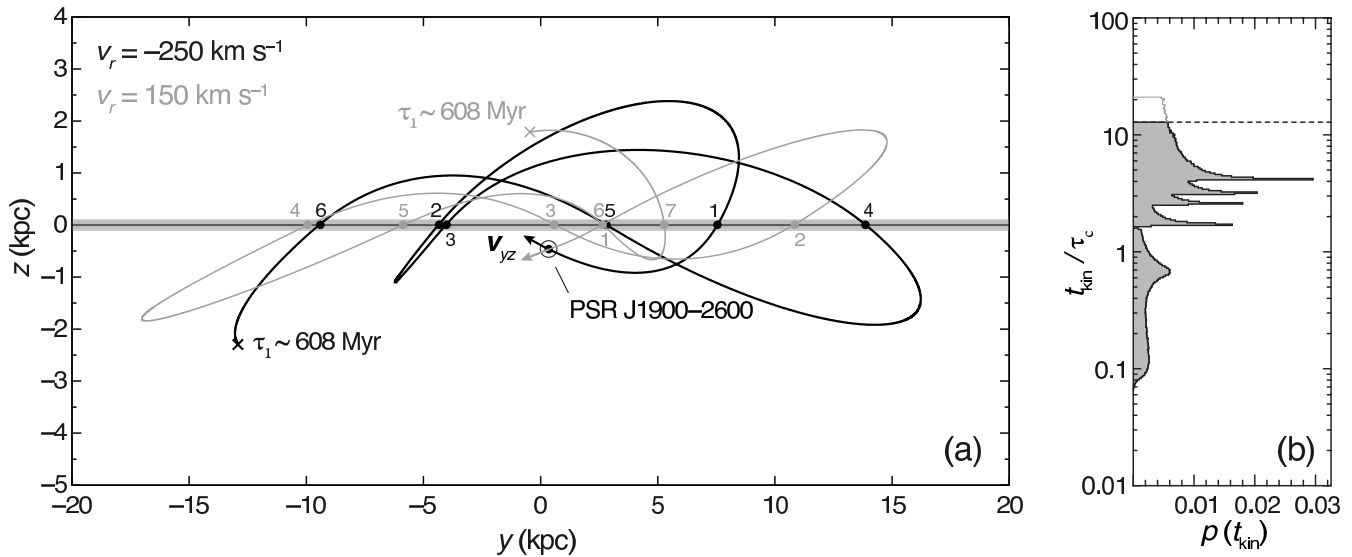} \caption{\label{fig:j1900traj} (a) Trajectory of PSR J1900$-$2600 between the pulsar's present position (circled dot) and that at $\tau_1 \sim 608$ Myr in the past (cross), assuming $v_r=-250$ km s$^{-1}$ (black curve) and $v_r=150$ km s$^{-1}$ (grey curve). The trajectories are projected on the $y$--$z$ plane that is perpendicular to the sightline between the Sun and the GC. The intersections with the GP along the two trajectories are numbered in ascending chronological order, the most recent being the first (1). (b) PDF of $t_{\rm kin}$ for PSR J1900$-$2600, as was derived from our kinematics simulation for $t_{\rm kin}\leq \tau_1$ (grey-filled histogram); for completeness, the histogram of solutions beyond $\tau_1$ (marked with a dashed line) is shown as a grey outline up to 1 Gyr.} 
\end{figure*}

\subsubsection{Multiple Intersections with the Galactic Plane} 
\label{sec:multisect} 

For the majority of pulsars in our sample, our simulation produces only a single intersection with the GP, within $\tau_1$ and for all investigated $v_r$ values. In contrast, Fig.~\ref{fig:j1900traj}a shows an example trajectory for PSR J1900$-$2600 (black line), between the pulsar's present position and that $\tau_1\sim 608$ Myr ago, if $v_r=-250$ km s$^{-1}$. Under these particular assumptions, it is seen that the pulsar crosses the GP 6 times, exiting the plane in between, which would thus produce 6 discrete solutions for $t_{\rm kin}$. The $p(t_{\rm kin})$ for this pulsar is shown in Fig.~\ref{fig:j1900traj}b, where one can see that it is composed of multiple peaks. However, it should be cautioned that the correspondence of the discrete peaks contained in $p(t_{\rm kin})$ to the crossings of the GP for this particular trajectory is not necessarily one-to-one, as the former PDF is the result of marginalisation over all $v_r$ values and, hence, over all possible trajectories; these may produce a different number of intersections and at different times. Here, we clarify that our simulation treated each possible solution for each trajectory up to $\tau_1$ as equally likely. Nevertheless, the different probabilities of each peak in the PDFs arise from the following: (a) as was stated above, certain $v_r$ values may lead to fewer number of intersections than others, which when marginalising over $v_r$ would lead to lower weighting at those $t_{\rm kin}$ values corresponding to the least occurring intersection number. This can be made clearer in Fig.~\ref{fig:j1900traj}a, where the trajectory for $v_r=-250$ km s$^{-1}$ (black line) is compared to that for $v_r=150$ km s$^{-1}$ (grey line). The latter produces an additional intersection, i.e.~7, within $\tau_1$, compared to the former case, which contributes additional solutions towards $p(t_{\rm kin})$ at roughly the same $t_{\rm kin}$ (note that $t_{\rm kin}$ at intersections 6 and 7 is roughly the same on a log scale --- see below). (b) In addition, for a single $v_r$, the crossing duration of an intersection (i.e.~the time a pulsar spends in $|z|\leq 0.1$ kpc), as well as the time interval between intersections affects the width and height of the peaks in $p(t_{\rm kin})$. In particular, if e.g.~two successive intersections occur close in time (i.e.~a small interval between peaks), then depending on the width of the peaks (i.e.~the crossing duration) the corresponding probability at a certain $t_{\rm kin}$ may be the sum of probabilities due to two or more intersections. And, of course, since $p(t_{\rm kin})$ is the average over all values of $v_r$, the shape of the peaks in the PDF is also determined by the spread in $t_{\rm kin}$ corresponding to each intersection, which is caused by the range of travel times for the different $v_r$ values.  
(c) Finally, the PDF of $t_{\rm kin}$ is presented on a log scale with constant bin size, $\Delta \log (t_{\rm kin}/{\rm Myr})=const$. As a consequence, uniformly distributed probability densities as a function of age on a linear scale would correspond to non-uniform probability densities on a logarithmic scale: this is simply expressed as $\Delta p(t_{\rm kin})/\Delta\log t_{\rm kin}\propto t_{\rm kin}\Delta p(t_{\rm kin})/\Delta t_{\rm kin}$. In other words, the number of solutions per age interval would statistically increase with $t_{\rm kin}$, favouring higher probabilities at older ages; however, this effect is mainly evident in PDFs having multiple peaks separated by at least an order of magnitude in $t_{\rm kin}$: as we explain in the following section, pulsars having such PDFs have been excluded from further analysis.

\begin{figure*} 
\vspace*{10pt} 
\includegraphics[width=1.0\textwidth]{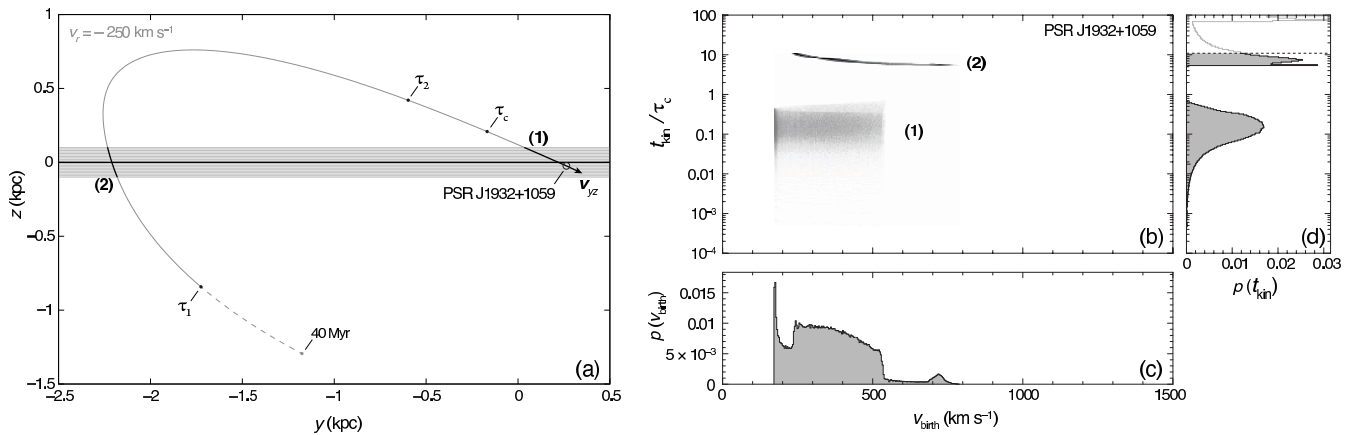} \caption{\label{fig:trajectories} (a) Grey curve: 40-Myr trajectory of PSR J1932+1059 for the case of $v_r=-250$ km s$^{-1}$, shown on the $y$--$z$ plane (see Section~\ref{subsec:method} for a definition). The pulsar's current position is shown with an open, black circle, where the $yz$ component of its present velocity assuming the above $v_r$ is also shown. The pulsar positions along the path at $\tau_{\rm c}$, $\tau_2$ and $\tau_1$ ago are also indicated with black bullets. Pulsar positions beyond $\tau_1$, here shown with a grey, dashed line, were not considered in our study. The horizontal, black line ($z=0$ kpc) and surrounding grey lines ($0<|z|\leq 0.1$ kpc) mark the range of possible $z_{\rm birth}$ values considered in our study; the range of possible birth sites of this pulsar, where the path shown intersects the different $z_{\rm birth}$ values, are highlighted as black, curved sections at positions (1) and (2). (b) Greyscale map of the 2D PDF of $v_{\rm birth}$ and $t_{\rm kin}$ derived from our MC simulation for PSR J1932+1059 (see Eq.~\ref{eq:2Dpdf}). The two separate regions of high probability in this PDF are the result of this pulsar having crossed the GP twice, an example of which is shown in (a). (c) Posterior PDF of $v_{\rm birth}$ derived from marginalising the 2D PDF over $t_{\rm kin}$ (see Eq.~\ref{eq:marginalPDFs}). (d) Posterior PDF of $t_{\rm kin}$ derived from marginalising the 2D PDF over $v_{\rm birth}$ (see Eq.~\ref{eq:marginalPDFs}). The horizontal, dashed line corresponds to $\tau_1$, i.e.~the upper limit of our simulation; the PDF for larger $t_{\rm kin}$ is shown for completeness as a grey outline.} 
\end{figure*}

\subsection{Results}
\label{subsec:results}
\subsubsection{The determined kinematic ages}
The majority of pulsars in our sample, i.e.~33 out of the 52, have a $p(t_{\rm kin})$ that exhibits a single, clear peak inside the investigated range (e.g.~Fig.~\ref{fig:pdfs}a). Such cases allow us to confidently assign the most probable kinematic age to each pulsar (i.e.~the peak value) and quote the associated 68\% upper and lower CLs around the peak. However, for the remaining 19 cases, the corresponding $p(t_{\rm kin})$ is complex, having multiple peaks --- in many cases, of comparable significance. An extreme case of this category is the PDF of PSR J0814+7429 (see Fig.~\ref{fig:pdfs}b), for which $\tau_1\sim 1.7$ Gyr, during which this pulsar's path has intersected with the GP more than 10 times (for certain values of $v_r$). In such cases, although we can assign a most probable $t_{\rm kin}$ to these pulsars based on the value at the highest peak, we cannot confidently exclude ages corresponding to other peaks of roughly equal probability to that of the highest. A few of those cases exhibit a clear bimodality in the shape of $p(t_{\rm kin})$, with a peak separation that can be up to 2 orders of magnitude in units of $\tau_{\rm c}$. Fig.~\ref{fig:trajectories} shows a characteristic example of this bimodality for the case of PSR J1932+1059, where $p(t_{\rm kin})$ contains two significant peaks near $t_{\rm kin}=0.1\tau_{\rm c}$ and $10\tau_{\rm c}$. The nature of those peaks becomes clearer in the grayscale map of $p(v_{\rm birth},t_{\rm kin})$, where two distinct regions of high probability are visible, each corresponding to the pulsar's intersection with the GP (i.e.~$|z|\leq 0.1$ kpc), separated by the interval when the pulsar was travelling outside the plane, i.e.~$|z|>0.1$ kpc (Fig.~\ref{fig:trajectories}b). The intersections are shown in Fig.~\ref{fig:trajectories}a, at positions (1) and (2), where we have chosen an example trajectory corresponding to $v_r=-250$ km s$^{-1}$. Therefore, when there is an ambiguity in $t_{\rm kin}$, we report only the value corresponding to the highest peak without attempting to estimate uncertainties. 

Table~\ref{tab:tab2} shows the resulting $t_{\rm kin}$ values from our MC analysis, for all pulsars with convergent solutions within $t_{\rm kin}\leq\tau_1$, $|v_r|<500$ km s$^{-1}$ and $|z|\leq 0.1$ kpc. All ambiguous determinations are denoted with an exclamation mark next to the value of $t_{\rm kin}$.

\begin{table*}
\renewcommand{\arraystretch}{1.3}
\caption{\label{tab:tab2}  
Age and kinematic properties of 52 pulsars (columns 4--9) --- taken from the ATNF catalogue (Manchester et al.~2005) --- for which we were able to determine a kinematic age, $t_{\rm kin}$ (column 11), in the interval 0--$\tau_1$ (column 10; see text for definition). The distance to these pulsars has been estimated using various methods, e.g.~from the pulsar DM and a model of the Galactic electron density, from VLBI observations, or from timing parallax. Columns 12--14 show the probability of having $t_{\rm kin}\leq \tau_{\rm c}$, $\tau_{\rm c}<t_{\rm kin}\leq \tau_2$ and $\tau_2<t_{\rm kin}\leq \tau_1$, respectively, in the considered age interval. The asymmetric errors on $t_{\rm kin}$ correspond to the 68\% CLs around the peak value. In several cases, the $t_{\rm kin}$ PDF contains multiple peaks and discontinuities: those cases, denoted with an exclamation mark, result in an ambiguous determination of $t_{\rm kin}$ and the peak value is reported without the corresponding CLs. For those pulsars for which the determination of $t_{\rm kin}$ was not ambiguous, the most probable value of the birth period for $n=3$, $P_{0,3}$, and its 68\% CLs is provided in the last column (see Section~\ref{sec:birthpers}).}
\small 
\centering
\begin{tabular}{@{}llrrrrrrrrrrrrr} 
\\
\hline 
$N$             &  PSR             &  $P$              & $\log\tau_{\rm c}$    & $l$         & $b$         & $d$         & $v_{l}$     & $v_{b}$  & $\log \tau_1$   &  $\log t_{\rm kin}$   & $p_3$ & $p_2$ & $p_1$  & $P_{0,3}$        \\
                &                  &  [ms]             &  [yr]                 &  [$^\circ$] & [$^\circ$]  & [kpc]       &  [km/s]     & [km/s]   & [yr]            &  [yr]                 &       &       &        & [ms] \\
\hline 
1      &    	J0139+5814      &          272    &                5.6    &   129.22    &  $-$4.04    &     2.89    & $-$249.7    & $-$111.2    & 6.7 &      $6.3_{-0.1}^{+0.1}$  &  0.00 & 0.00 & {\bf 1.00}    &  -- \\
2      &    	J0152$-$1637    &          833    &                7.0    &   179.31    & $-$72.46    &     0.79    &     22.6    &  $-$61.6    & 8.1 &      $8.0$ \ (!)             &  0.42 & 0.04 & {\bf 0.54} &    --        \\
3      &    	J0304+1932      &         1388    &                7.2    &   161.14    & $-$33.27    &     0.95    &     90.6    & $-$113.4    & 8.4 &      $8.3$ \ (!)             &  0.27 & 0.02 & {\bf 0.71}  &    --    \\
4      &    	J0332+5434      &          714    &                6.7    &   145.00    &  $-$1.22    &     1.06    &     91.5    &     17.4    & 7.9 &      $7.4$ \ (!)             &  0.20 & 0.00 & {\bf 0.80}  &    --  \\
5      &    	J0358+5413      &          156    &                5.8    &   148.19    &     0.81    &     1.10    &      1.9    &     70.4    & 6.8 &      $5.7_{-0.5}^{+0.3}$  &  {\bf 0.65} & 0.28 & 0.07    &  $128_{-47}^{+18}$ \\
6      &    	J0452$-$1759    &          549    &                6.2    &   217.08    & $-$34.09    &     3.14    & $-$141.6    &    245.6    & 7.3 &      $7.3$ \ \ \ \ \                 &  0.00 & 0.00 & {\bf 1.00} &  -- \\
7      &    	J0454+5543      &          341    &                6.4    &   152.62    &     7.55    &     0.79    &    160.1    &    119.1    & 7.4 &      $5.9_{-0.2}^{+0.2}$  &  {\bf 0.98} & 0.02 & 0.00   & $291_{-55}^{+22}$\\
8      &    	J0538+2817      &          143    &                5.8    &   179.72    &  $-$1.69    &     1.47    & $-$408.1    &     63.7    & 6.8 &      $5.8_{-0.6}^{+0.1}$  & {\bf 0.87}  & 0.13 & 0.00   & $141_{-48}^{+1}$ \\
9      &    	J0630$-$2834    &         1244    &                6.4    &   236.95    & $-$16.76    &     2.15    & $-$313.6    & $-$348.4    & 7.6 &      $6.2_{-0.1}^{+0.2}$  & {\bf 0.88}  & 0.12 & 0.00  & $848_{-268}^{+51}$ \\
10     &    	J0659+1414      &          385    &                5.0    &   201.11    &     8.26    &     0.29    &     23.7    &     62.1    & 6.1 &      $5.7_{-0.4}^{+0.3}$  & 0.07  & 0.09 & {\bf 0.83}  & $383_{-177}^{+1}$ \\
11     &    	J0738$-$4042    &          375    &                6.6    &   254.19    &  $-$9.19    &    11.03    & $-$798.3    & $-$337.6    & 7.6 &      $6.7^{+0.1}$         & 0.00  & {\bf 1.00} & 0.00  & -- \\
12     &    	J0742$-$2822    &          167    &                5.2    &   243.77    &  $-$2.44    &     1.89    & $-$121.8    & $-$202.6    & 6.2 &      $5.6_{-0.3}^{+0.2}$  & 0.10  & 0.20 & {\bf 0.71}   & $139_{-67}^{+19}$ \\
13     &    	J0814+7429      &         1292    &                8.1    &   140.00    &    31.62    &     0.43    &     57.1    &     70.8    & 9.2 &      $7.9$ \ (!)            & {\bf 0.46}  & 0.19 & 0.35  & -- \\ 
14     &    	J0826+2637      &          531    &                6.7    &   196.96    &    31.74    &     0.36    &    145.0    &     68.6    & 7.8 &      $7.8$ \ (!)             & {\bf 0.79}  & 0.06 & 0.16  & -- \\
15     &    	J0835$-$4510    &         89.3    &                4.1    &   263.55    &  $-$2.79    &     0.29    &  $-$58.2    &  $-$23.7    & 5.0 &      $5.0$ \ (!)            & 0.11  & 0.11 & {\bf 0.78}  & -- \\
16     &    	J0837+0610      &         1274    &                6.5    &   219.72    &    26.27    &     0.72    & $-$132.3    &    101.7    & 7.6 &      $6.1_{-0.1}^{+0.5}$  & {\bf 0.66}  & 0.17 & 0.17 & $995_{-262}^{+49}$ \\
17     &    	J0837$-$4135    &          752    &                6.5    &   260.90    &  $-$0.34    &     4.24    &    357.0    & $-$249.1    & 7.6 &      $5.3_{-0.5}^{+0.2}$  & {\bf 1.00}  & 0.00 & 0.00  & $740_{-19}^{+7}$ \\
18     &    	J0922+0638      &          431    &                5.7    &   225.42    &    36.39    &     1.20    & $-$296.0    &    353.0    & 6.8 &      $6.2_{-0.1}^{+0.3}$  & 0.00  & 0.00 & {\bf 1.00}  & -- \\
19     &    	J0953+0755      &          253    &                7.2    &   228.91    &    43.70    &     0.26    &  $-$21.6    &     32.3    & 8.3 &      $5.8$ \ (!)             & 0.44  & 0.02 & {\bf 0.55}  & -- \\
20     &    	J1136+1551      &         1188    &                6.7    &   241.90    &    69.20    &     0.36    & $-$205.3    &    283.5    & 7.9 &      $6.0_{-0.2}^{+0.6}$  & {\bf 0.86}  & 0.06 & 0.08   & $1097_{-142}^{+18}$ \\
21     &    	J1239+2453      &         1382    &                7.4    &   252.45    &    86.54    &     0.86    &  $-$25.9    & $-$177.4    & 8.5 &      $7.6$ \ (!)             & {\bf 0.61}  & 0.16 & 0.23  &  -- \\
22     &    	J1430$-$6623    &          785    &                6.7    &   312.65    &  $-$5.40    &     1.80    & $-$265.6    &  $-$57.9    & 7.8 &      $7.4$ \ (!)             & 0.32  & 0.12 & {\bf 0.56}  & -- \\
23     &    	J1453$-$6413    &          180    &                6.0    &   315.73    &  $-$4.43    &     1.84    & $-$164.3    &  $-$93.6    & 7.0 &      $6.1_{-0.2}^{+0.2}$  & 0.15  & {\bf 0.59} & 0.26  & $77_{-36}^{+38}$ \\
24     &    	J1456$-$6843    &          263    &                7.6    &   313.87    &  $-$8.54    &     0.45    &  $-$65.9    &     23.5    & 8.7 &      $7.8_{-0.4}^{+0.4}$  & 0.19  &  0.3 & {\bf 0.51}  & $263_{-127}^{+1}$ \\
25     &    	J1509+5531      &          740    &                6.4    &    91.33    &    52.29    &     2.41    &  $-$51.7    &   1091.8    & 7.5 &      $6.4_{-0.1}^{+0.2}$  & 0.35  & {\bf 0.51} & 0.14  & $263_{-107}^{+84}$ \\
26     &    	J1604$-$4909    &          327    &                6.7    &   332.15    &     2.44    &     3.59    & $-$338.5    &    331.2    & 7.8 &      $5.6_{-0.1}^{+0.1}$  & {\bf 1.00}  & 0.00 & 0.00  & $313_{-4}^{+4}$ \\
27     &    	J1645$-$0317    &          388    &                6.5    &    14.11    &    26.06    &     2.91    &    302.8    &    276.5    & 7.6 &      $6.6_{-0.1}^{+0.3}$  & 0.27  & {\bf 0.44} & 0.28  & $169_{-67}^{+19}$  \\
28     &    	J1709$-$1640    &          653    &                6.2    &     5.77    &    13.66    &     1.27    &     19.5    &   $-$9.7    & 7.3 &      $6.5_{-0.1}^{+0.5}$  & 0.00  & 0.22 & {\bf 0.78}  & -- \\
29     &    	J1735$-$0724    &          419    &                6.7    &    17.27    &    13.28    &     4.32    &    490.7    &    347.9    & 7.8 &      $6.4_{-0.1}^{+0.1}$  & {\bf 1.00}  & 0.00 & 0.00  & $309_{-46}^{+16}$ \\
30     &    	J1740+1311      &          803    &                6.9    &    37.08    &    21.68    &     4.77    & $-$491.1    &    294.8    & 8.0 &      $6.7_{-0.1}^{+0.3}$  & {\bf 0.78}  & 0.22 & 0.00  &  $568_{-176}^{+33}$ \\
31     &    	J1801$-$2451    &          125    &                4.2    &     5.25    &  $-$0.88    &     4.61    &  $-$22.2    &  $-$63.8    & 5.2 &      $5.2_{-0.5}$         & 0.09  & 0.10 & {\bf 0.81}  & $124_{-57}$ \\
32     &    	J1820$-$0427    &          598    &                6.2    &    25.46    &     4.73    &     2.45    &   $-$6.7    &    147.7    & 7.3 &      $7.2$ \ (!)             & {\bf 0.64}  & 0.23 & 0.14  & -- \\
33     &    	J1844+1454      &          376    &                6.5    &    45.56    &     8.15    &     2.23    &    422.0    &    303.6    & 7.6 &      $6.0_{-0.1}^{+0.1}$  & {\bf 1.00}  & 0.00 & 0.00  & $314_{-17}^{+10}$ \\

\hline 
\end{tabular} 

\end{table*}

\setcounter{table}{0}

\begin{table*}
\renewcommand{\arraystretch}{1.3}
\caption{\label{tab:tab2}  Continued.} 
\small
\centering 
\begin{tabular}{@{}llrrrrrrrrrrrrr} 
\\
\hline 
$N$             &  PSR             &  $P$              & $\log\tau_{\rm c}$    & $l$         & $b$         & $d$         & $v_{l}$     & $v_{b}$  & $\log \tau_1$      &  $\log t_{\rm kin}$  & $p_3$ & $p_2$ & $p_1$   & $P_{0,3}$  \\
                &                  &  [ms]             &  [yr]                 &  [$^\circ$] & [$^\circ$]  & [kpc]       &  [km/s]     & [km/s]   & [yr]               &  [yr]                &       &       &         & [ms]         \\
\hline 
34     &    	J1850+1335      &          346    &                6.6    &    44.99    &     6.34    &     3.14    &    249.3    &    141.2    & 7.6 &      $6.3$ \ (!)            & {\bf 0.70}  & 0.02  & 0.29 & -- \\
35     &    	J1900$-$2600    &          612    &                7.7    &    10.34    & $-$13.45    &     2.00    & $-$462.6    &   $-$2.6    & 8.8 &      $8.3$ \ (!)           & 0.27  & 0.08  & {\bf 0.64}  & -- \\
36     &    	J1907+4002      &         1236    &                7.6    &    70.95    &    14.20    &     1.76    &    154.7    &  $-$41.2    & 8.7 &      $7.4$ \ (!)           & 0.30  & 0.04  & {\bf 0.66}  & --  \\
37     &    	J1913$-$0440    &          826    &                6.5    &    31.31    &  $-$7.12    &     3.22    &     23.9    & $-$127.4    & 7.6 &      $7.2$ \ (!)            & {\bf 0.36}  & 0.32  & 0.33  & -- \\
38     &    	J1915+1009      &          404    &                5.6    &    44.71    &  $-$0.65    &     5.32    &    204.1    & $-$156.2    & 6.7 &      $5.6_{-0.3}^{+0.2}$ & {\bf 0.50}  & 0.43  & 0.07  & $247_{-117}^{+99}$  \\
39     &    	J1921+2153      &         1337    &                7.2    &    55.78    &     3.50    &     0.66    &    124.2    &      6.4    & 8.4 &      $7.6$ \ (!)           & 0.27  & 0.18  & {\bf 0.55} & -- \\
40     &    	J1932+1059      &          226    &                6.5    &    47.38    &  $-$3.88    &     0.36    &    146.8    &  $-$98.4    & 7.5 &      $7.2$ \ (!)            & {\bf 0.69}  & 0.00  & 0.31  &   --  \\
41     &    	J1935+1616      &          359    &                6.0    &    52.44    &  $-$2.09    &     7.93    & $-$224.3    & $-$198.5    & 7.0 &      $6.2_{-0.1}^{+0.1}$ & 0.00  & {\bf 0.96}  & 0.04  & $63_{-30}^{+15}$ \\
42     &    	J1937+2544      &          201    &                6.7    &    60.84    &     2.27    &     2.76    &    256.9    &  $-$91.5    & 7.7 &      $7.3^{+0.3}$        & 0.00  & 0.00  & {\bf 1.00}  & -- \\
43     &    	J1952+3252      &         39.5    &                5.0    &    68.77    &     2.82    &     2.50    & $-$166.6    &    202.8    & 5.9 &      $5.8_{-0.2}^{+0.1}$ & 0.00  & 0.05  & {\bf 0.95}  & $7_{-3}^{+3}$ \\
44     &    	J1955+5059      &          519    &                6.8    &    84.79    &    11.55    &     1.80    &    338.5    &    400.5    & 7.9 &      $5.9_{-0.1}^{+0.1}$ & {\bf 1.00}  & 0.00  & 0.00 & $482_{-10}^{+6}$ \\
45     &    	J2018+2839      &          558    &                7.8    &    68.10    &  $-$3.98    &     0.97    &  $-$11.5    &      1.2    & 8.9 &      $7.4$ \ (!)           & {\bf 0.45}  & 0.25  & 0.30 &  -- \\
46     &    	J2022+2854      &          343    &                6.5    &    68.86    &  $-$4.67    &     2.70    & $-$214.3    & $-$118.5    & 7.5 &      $6.3_{-0.1}^{+0.2}$ & {\bf 0.91}  & 0.09  & 0.00  & $231_{-72}^{+25}$ \\
47     &    	J2022+5154      &          529    &                6.4    &    87.86    &     8.38    &     2.00    &    104.4    &    106.8    & 7.5 &      $6.3_{-0.1}^{+0.3}$ & {\bf 0.57}  & 0.37  & 0.06  & $325_{-116}^{+37}$ \\
48     &    	J2048$-$1616    &         1962    &                6.5    &    30.51    & $-$33.08    &     0.64    &    107.4    & $-$325.3    & 7.6 &      $5.9$ \ (!)            & {\bf 0.60}  & 0.09  & 0.30  & -- \\
49     &    	J2157+4017      &         1525    &                6.8    &    90.49    & $-$11.34    &     5.58    &    515.4    & $-$221.1    & 8.0 &      $6.6_{-0.1}^{+0.2}$ & {\bf 0.82}  & 0.18  & 0.00   & $968_{-327}^{+89}$ \\
50     &    	J2219+4754      &          538    &                6.5    &    98.38    &  $-$7.60    &     2.45    & $-$259.8    & $-$203.7    & 7.6 &      $6.2_{-0.1}^{+0.1}$ & {\bf 1.00}  & 0.00  & 0.00  & $397_{-76}^{+24}$ \\
51     &    	J2305+3100      &         1576    &                6.9    &    97.72    & $-$26.66    &     3.92    &  $-$56.5    & $-$328.1    & 8.1 &      $6.6_{-0.1}^{+0.4}$ & {\bf 0.71}  & 0.23  & 0.06  & $1176_{-334}^{+40}$ \\
52     &    	J2330$-$2005    &         1644    &                6.7    &    49.39    & $-$70.19    &     0.49    &     29.0    & $-$139.5    & 7.9 &      $6.0$ \ (!)            & 0.46  &  0.05 & {\bf 0.49} &  -- \\

\hline 
\end{tabular} 

\footnotesize \flushleft (!) Complex PDF with multiple significant peaks and in some cases discontinuities. Only the value corresponding to the PDF maximum is tabulated.  \\ \normalsize

\end{table*}

\subsubsection{The reliable sample of kinematic ages}
Despite the difficulties in assigning a single, most-likely kinematic age to several pulsars in our sample, it is still useful for all pulsars to report how probable it is that the true age --- based on kinematics and all our assumptions --- lies within the intervals demarcated by the characteristic time scales $\tau_{\rm c}$, $\tau_2$ and $\tau_1$. These probabilities were calculated for each pulsar as 
\begin{align}
\nonumber
p_3 &= p(t_{\rm kin}\leq\tau_{\rm c}) = \int_0^{\tau_{\rm c}}p(t_{\rm kin})dt_{\rm kin}\\
p_2 &= p(\tau_{\rm c}<t_{\rm kin}\leq\tau_2) = \int_{\tau_{\rm c}}^{\tau_2}p(t_{\rm kin})dt_{\rm kin}\\
\nonumber
p_1 &= p(\tau_2<t_{\rm kin}\leq\tau_1) = \int_{\tau_2}^{\infty}p(t_{\rm kin})dt_{\rm kin}
\end{align}
and are shown in columns 12--14 of Table~\ref{tab:tab2}, with the highest probability of the three shown in boldface.

In addition, the probabilities $p_1, p_2$ and $p_3$ can be used to discuss the most-probable range of braking indices for our pulsar sample. In Section~\ref{sec:conclusions}, we investigate the implications of the $t_{\rm kin}$ distributions for the distribution of $n$.

Our analysis was restricted to $t_{\rm kin}\leq \tau_1$. However, given enough time, many of the pulsars in our sample would intersect with the GP beyond $\tau_1$, effectively providing an infinite number of solutions for $t_{\rm kin}$. For example, the case of PSR J1932+1059 (Fig.~\ref{fig:trajectories}) clearly demonstrates that allowing only $t_{\rm kin}\leq \tau_1$ truncates $p(t_{\rm kin})$ near the position of the most probable value of $t_{\rm kin}$. It is also clear from that plot that the PDF exhibits another peak of even higher probability at $t_{\rm kin}\sim 100\tau_{\rm c}$ or $10\tau_1$ (shown with a grey line). Nevertheless, such large departures from the characteristic timescales for this pulsar are beyond the scope of the present investigation.

After excluding all pulsars that did not satisfy the above criteria --- those marked with an exclamation mark in Table~\ref{tab:tab2} --- we were left with 33 pulsars for which $t_{\rm kin}$ was considered reliable for the present work. All the derived $p(t_{\rm kin})$ for the selected sample are shown in Fig.~\ref{fig:selectPDFs}. Using these pulsars alone, we wish to re-examine the distributions of Fig.~\ref{fig:fig5}, where the age bins are now defined by the `most probable $t_{\rm kin}$' instead of $\tau_{\rm c}$. Hereafter, when we mention the $t_{\rm kin}$ value of a pulsar, we are referring to the most probable value in the corresponding $p(t_{\rm kin})$. A comparison between the distribution of $\tau_{\rm c}$, using all 58 PSRs, and that of $t_{\rm kin}$, using the MC values produced by our simulation for the 33 PSRs that passed our criteria, is shown in Fig.~\ref{fig:ageDists}. The median value for both distributions is also shown in the same figure, with \textless$\log (t_{\rm kin}/{\rm yr})$\textgreater$\,=6.2_{-0.6}^{+0.5}$ and \textless$\log (\tau_{\rm c}/{\rm yr})$\textgreater$\,=6.5_{-1.1}^{+0.5}$. Although the PDF of $t_{\rm kin}$ appears to be slightly shifted towards younger pulsar ages compared to the PDF of $\tau_{\rm c}$, the difference is not significant compared to the $1\sigma$ confidence intervals of the distributions. 

\subsubsection{Very young pulsars}
\begin{figure} \vspace*{10pt} \includegraphics[width=0.48\textwidth]{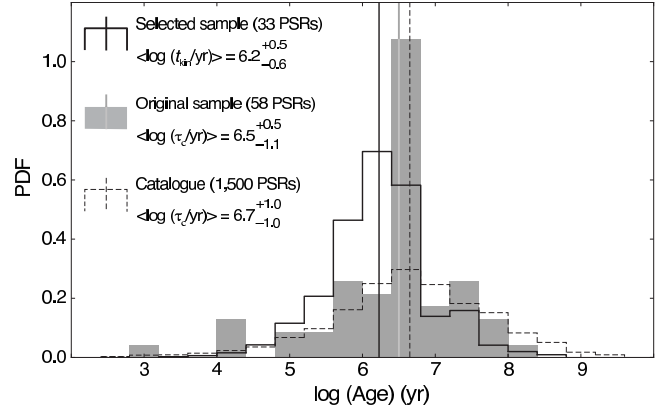} \caption{\label{fig:ageDists} Grey-filled bins: $\tau_{\rm c}$ PDF of the original sample of 58 pulsars considered in the herein study of spin--velocity alignment. Solid-line steps: $t_{\rm kin}$ PDF generated from all MC values produced by our simulation for 33 pulsars with a reliable $t_{\rm kin}$ determination (see Table~\ref{tab:tab2}). Dashed-line steps: $\tau_{\rm c}$ PDF of 1,500 non-recycled pulsars from PSRCAT. The vertical lines of corresponding line style indicate the median of each distribution. The medians and the 68\% CLs are also shown in the inset key.} 
\end{figure}

\begin{figure} 
\vspace*{10pt} 
\includegraphics[width=0.48\textwidth]{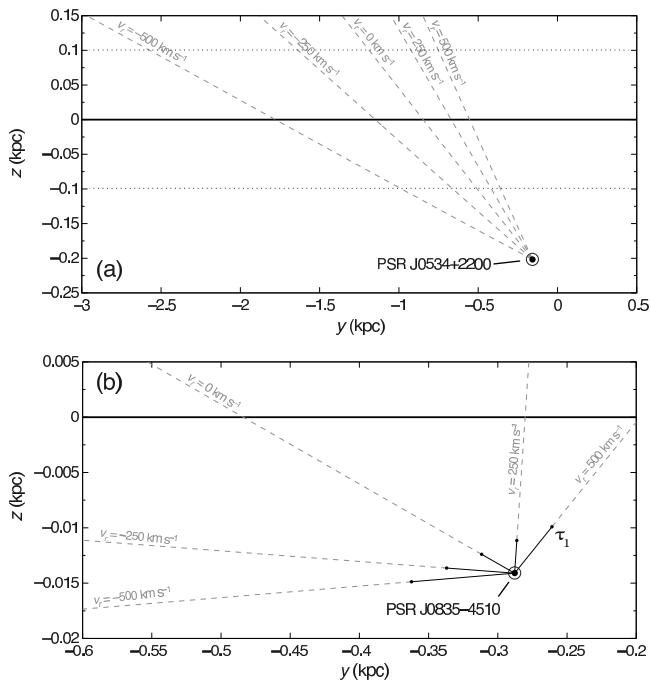} \caption{\label{fig:crabvela} Example trajectories of the (a) Crab and (b) Vela pulsars, on the $y$--$z$ plane, between their present positions (circled dots) and those in the past, delineated with grey dashed lines, for various magnitudes of the radial velocity, $v_r$. Our simulation only considered path lengths of up to $\tau_1$ in the past (black solid lines); for the Crab pulsar, $\tau_1$ is very short and cannot be seen on the chosen scale. The investigated range of birth heights in our work was $|z_{\rm birth}|\leq 100$ pc, which is demarcated with black, horizontal, dotted lines.} 
\end{figure}

A notable difference between the distribution of $t_{\rm kin}$ for the selected 33 pulsars and that of $\tau_{\rm c}$ for the original sample of 58 pulsars (Fig.~\ref{fig:ageDists}) is the longer low-age tail of the latter, extending below 100 kyr. This is largely due to the characteristic ages of the Crab and Vela pulsars. The Crab pulsar is missing from the sample of $t_{\rm kin}$ because within $\tau_1=2\tau_{\rm c}^{\rm Crab}\ln (P_{\rm Crab}/{\rm 1 \ ms})\approx 9$ kyr, where $\tau_{\rm c}^{\rm Crab}=1,240$ y, it never reaches within 100 pc of the GP. As a result, our simulation did not produce any solutions for this pulsar (see Fig.~\ref{fig:crabvela}a). The Vela pulsar ($\tau_1^{\rm Vela}\approx 100$ kyr) is excluded on the same grounds (see Fig.~\ref{fig:crabvela}b): it is presently found in the plane, but its motion in the $z$ direction is very limited within $\tau_1^{\rm Vela}$ to result in a reliable $t_{\rm kin}$.


Amongst the pulsars with a reliable $t_{\rm kin}$ determination are PSRs J0538+2817, J1801$-$2451 and J1952+3252, whose ages have been previously estimated from either pulsar timing and/or VLBI measurements of their outwards proper motion from the respective centres of their parent SNRs (Kramer et al.~2003\nocite{klh+03}; Ng et al.~2007\nocite{nrb+07}; Gaensler et al.~2000\nocite{gf00}; Zeiger et al.~2008\nocite{zbcg08}). From that work, the true age of PSR J0538+2817 was determined to be $20$--$40$ kyr, roughly an order of magnitude younger than its characteristic age. Our simulations assign a $t_{\rm kin} = 631_{-473}^{+164}$ kyr, for this pulsar, with a probability of $p(t_{\rm kin}<40 \ {\rm kyr})\approx 7$\%. Furthermore, the true age of PSR J1801$-$2451 is quite uncertain and could be as old as 170 kyr, if it was born in G5.4$-$1.2, or as young as $\tau_{\rm c}=15.5$ kyr, if the association is false and the characteristic age is close to the true age. Our simulation yields $t_{\rm kin}\leq 158$ kyr, with a 92\% probability that $t_{\rm kin}>\tau_{\rm c}$, which is consistent with the above range of estimates. Lastly, PSR J1952+3252 has been firmly associated with CTB 80 which implies a true age of $\sim 51$ kyr in contrast to its $\sim 100$-kyr characteristic age. Our simulation does not predict such a young age, producing instead $t_{\rm kin}=631_{-233}^{+163}$ kyr, with $p(t_{\rm kin}\leq 96 \ {\rm kyr})=0$. Similarly to the cases of the Crab and Vela pulsars, the above comparisons highlight the weakness of our kinematic method in predicting the ages of very young pulsars, which have not traversed significant lengths through the Galaxy, away from their birth locations. In Section~\ref{subsubsec:vyp}, we explore the impact of these young pulsars on the overall trend of spin--velocity correlation with pulsar age, by considering their true ages.
     
In conclusion, our selection based on the degree of ambiguity in the determination of $t_{\rm kin}$ results in a somewhat narrower distribution of ages than that of $\tau_{\rm c}$, which is also quantitatively revealed by comparing the variances of the two distributions: ${\rm Var}(\tau_{\rm c})=0.9$ and ${\rm Var}(t_{\rm kin})=0.4$. This difference is largely due to the two youngest pulsars missing from the selected sample: the Crab and Vela pulsars. However, the selected sample is representative of kinematic ages covering a wide range of pulsar ages, ranging from $\sim 0.1$ to 100 Myr; this allows us to use the derived $t_{\rm kin}$ values as an independent variable in the investigation of spin--velocity alignment as a function of pulsar age, like we did in Section~\ref{sec:datanal} with $\tau_{\rm c}$.

\subsection{Spin--Velocity Alignment Revisited}
\label{subsec:spvrev} 
The $p(t_{\rm kin})$ derived in this work provide a quantitative tool for assigning an age to each pulsar, at a certain confidence level, that is an alternative to the characteristic age. Hence, it is also interesting to re-examine the spin--velocity alignment as a function of pulsar age, as was done in Fig.~\ref{fig:ksprobs1} (see Section~\ref{sec:datanal}), in the light of the newly derived kinematic ages for the pulsars in our sample. As was mentioned earlier, our re-investigation of the spin--velocity alignment included only the 33 pulsars of Fig.~\ref{fig:selectPDFs}. Similarly to Fig.~\ref{fig:fig5}, the tabulated distributions of $\Psi$, for intervals of $t_{\rm kin}$ instead of $\tau_{\rm c}$, are shown in Fig.~\ref{fig:fig5alt}. As before, in this plot each cell contains the distribution of $\Psi$ generated from pulsars in intervals of $t_{\rm kin}$, ranging from 100 to 100,000 kyr; also, the corresponding KS probabilities of rejecting uniformity are shown with each distribution. For comparison, we have overlaid the corresponding distributions of Fig.~\ref{fig:fig5} with dashed lines. Similarly to Fig.~\ref{fig:ksprobs1}, a scatter plot of the degree of spin--velocity correlation as a function of $t_{\rm kin}$ is shown in Fig.~\ref{fig:ksprobs2}. 

The direct comparison between $p_{\rm KS}(t_{\rm kin})$ and $p_{\rm KS}(\tau_{\rm c})$ reveals a very similar behaviour for pulsar ages $<10$ Myr, with both distributions rejecting that spin and velocity are uncorrelated with $\gtrsim 80\%$ confidence. There are indeed small differences between the central values of $p_{\rm KS}(t_{\rm kin})$ and $p_{\rm KS}(\tau_{\rm c})$, but these could be partly attributed to the different sample sizes; furthermore, if we take into account the 68\% CLs of $p_{\rm KS}$, both cases are consistent with each other. However, above 10 Myr, which roughly equates to $t_{\rm dyn}$, the PDF of $\Psi$ is evidently more uniform compared to those for younger pulsars; this is also confirmed by the small value of $p_{\rm KS}(t_{\rm kin})\approx 0.34$. The flattening of the $\Psi$ distribution for samples containing older pulsars is more clearly displayed in the scatter plot of Fig.~\ref{fig:ksprobs2}, where a downwards trend for $p_{\rm KS}$ can already be seen across the $1$--$10$ Myr range --- although the uncertainties on $p_{\rm KS}$ are quite significant. The $\Psi$ distribution for pulsars with $10 < t_{\rm kin} < 100$ Myr is evidently more uniform than those based on younger pulsars, which is confirmed by the low value of $p_{\rm KS}$. However, we should warn that the bin corresponding to the oldest kinematic ages contains only 3 pulsars, which is just below the margin for a reliable report of the KS statistic (Stephens 1970\nocite{ste70}). Interestingly, as can be seen in the same plot, spin--velocity alignment is retained at a high confidence ($p_{\rm KS}(\tau_{\rm c}>10 \ Myr)\approx 98\%$) for pulsars in the same age interval according to $\tau_{\rm c}$. Nonetheless, it should be noted that of the 12 pulsars with $\tau_{\rm c}\geq 10$ Myr in the original sample, only PSR J1456$-$6843 has a reliable $t_{\rm kin}$ in the corresponding age interval; the rest of the pulsars have complex $p(t_{\rm kin})$ and were therefore rejected. Two pulsars, PSRs J0452$-$1759 and J1937+2544, have a $t_{\rm kin}$ that is roughly an order of magnitude older than their characteristic ages $\tau_{\rm c}<10$ Myr; these are included in the top age interval of the new table. Lastly, since none of the pulsars in our sample have $t_{\rm kin}\gg 100$ Myr, we are not able to exclude spin--velocity correlation for very old pulsars. However, given the observed trend of $p_{\rm KS}$ with $t_{\rm kin}$, we consider it very likely that this is the case.

In conclusion, our alternative investigation of spin--velocity alignment as a function of pulsar age, based on pulsar kinematic ages, resulted in an intriguing flattening of the $\Psi$ distribution beyond 10 Myr. Although only 3 pulsars have $t_{\rm kin}>10$ Myr, the overall downward trend in $p_{\rm KS}$ for $t_{\rm kin}<10$ Myr can indicate that the Galactic gravitational potential plays indeed an important role in smearing out the observed spin--velocity alignment for older pulsars. 

\subsubsection{Very young pulsars}
\label{subsubsec:vyp}
\begin{figure*} 
\vspace*{10pt} 
\includegraphics[width=1\textwidth]{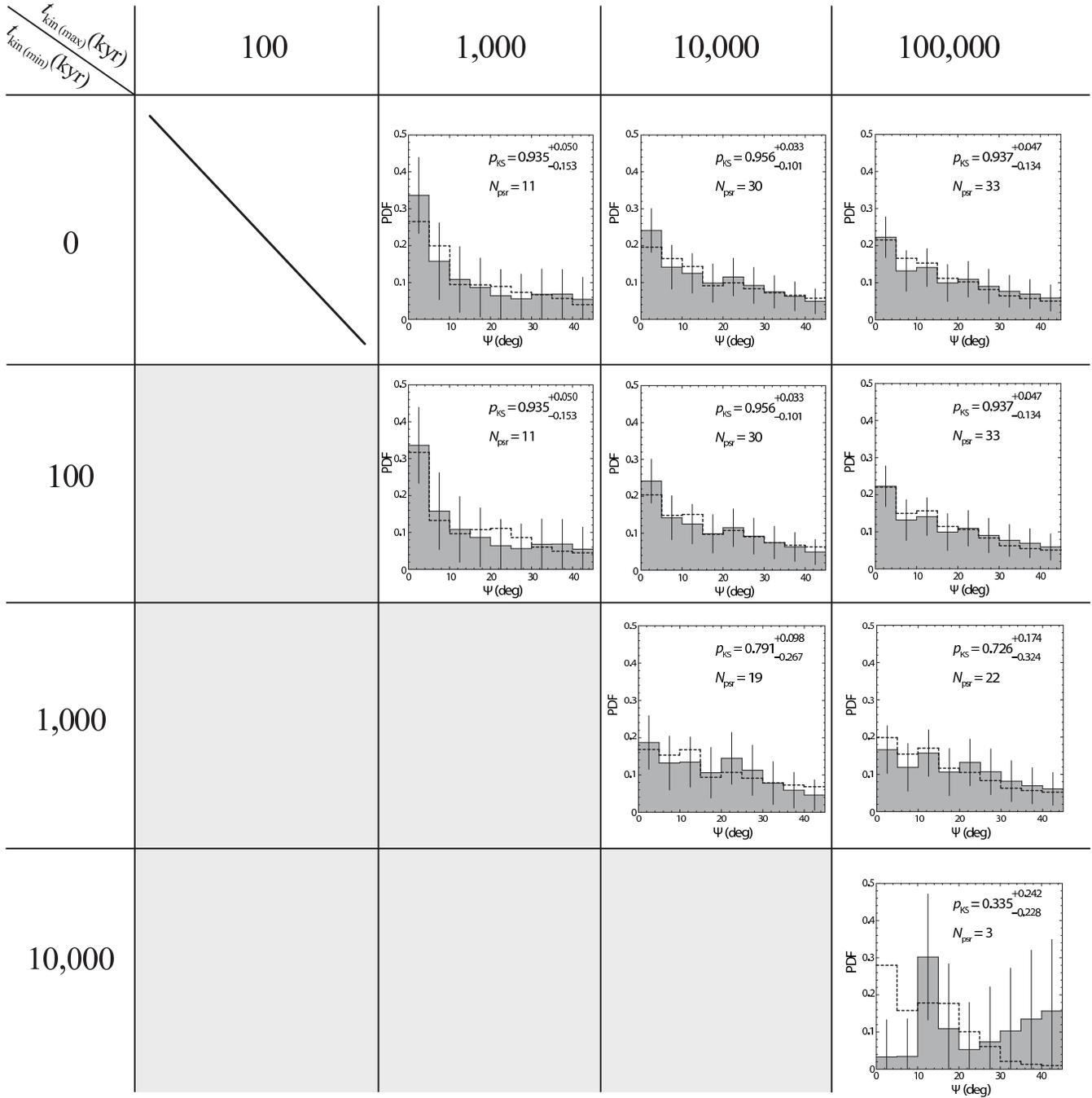} \caption{\label{fig:fig5alt} Tabulated distributions of $\Psi$ in intervals of $t_{\rm kin}$, for 33 pulsars of the original sample of Fig.~\ref{fig:fig5}, after omitting 19 pulsars with poor $t_{\rm kin}$ determination. The details of the statistical analysis are identical with those for Fig.~\ref{fig:fig5} (see corresponding caption). The $\Psi$ distributions shown here can be directly compared with those of Fig.~\ref{fig:fig5}, for the corresponding intervals in $\tau_{\rm c}$: these are shown with dashed, red lines on the same scale. Our analysis did not produce reliable kinematic ages below 100 kyr, so the corresponding interval in this plot is empty. A discussion on pulsars with very young ages can be found in Sections~\ref{subsec:results} and \ref{subsec:spvrev}.} 
\end{figure*}

There are a number of young pulsars in our sample for which we have independent estimates of their true ages, based on their SNR associations. Furthermore, the spin orientations of most of these young pulsars have been derived from fits to pulsar wind nebulae (PWNe) seen in their X-ray images (see Table~2 of Ng \& Romani 2004\nocite{nr04}). So, given the well-measured orientation of these pulsars, it is interesting to discuss their impact on the overall picture of Fig.~\ref{fig:fig5alt} by considering their true ages.

In the previous section, we mentioned the age estimates for PSRs J0538+2817, J1801$-$2451 and J1952+3252 from VLBI measurements of their outwards proper motion in their SNRs. In addition, the Crab and Vela pulsars, for which we were not able to derive a reliable $t_{\rm kin}$, have well-known true ages: the Crab pulsar has a historical age of 958 y, whereas the outwards proper motion of the Vela pulsar from the centre of its SNR implies an age of $18\pm 9$ kyr (Aschenbach et al.~1995\nocite{aet95}). Moreover, the age of the Vela pulsar has been independently estimated to be 20--30 kyr, from measurements of the pulsar's braking index ($n\approx 1.4$) and assuming a birth period of $P_0\approx 20$ ms (Lyne et al.~1996\nocite{lpgc96}). Also, the spin directions of these young pulsars from the PWNe fits result in small spin--velocity alignment angles, with $\Psi\sim 5^\circ$--$12^\circ$. The exception is PSR J1801$-$2451, for which the spin-axis orientation has been derived from polarisation measurements and results in $\Psi=35^\circ\pm 5^\circ$ (Johnston et al.~2007\nocite{jkk+07}). 

\begin{figure} 
\vspace*{10pt} 
\includegraphics[width=0.48\textwidth]{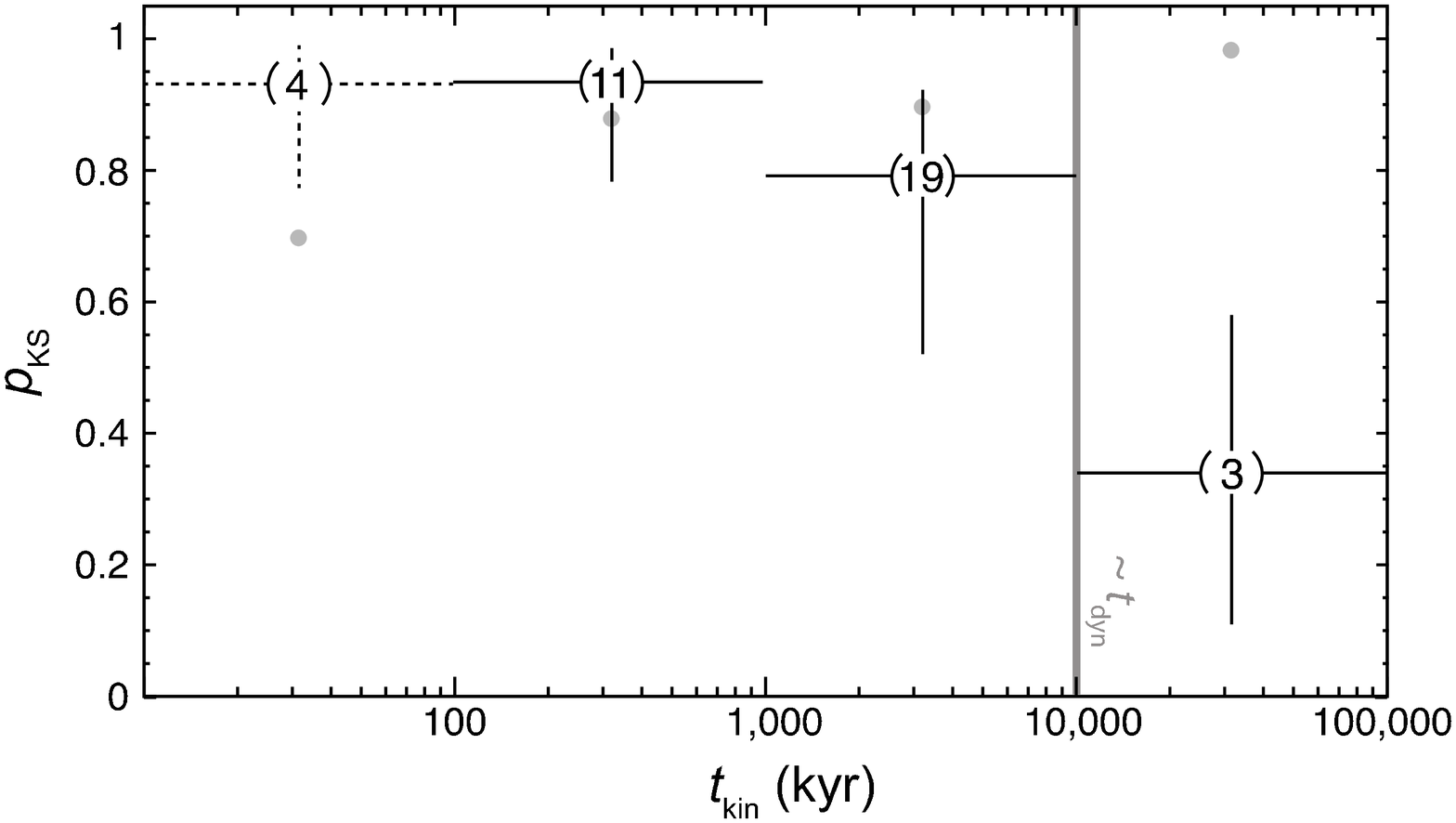} \caption{\label{fig:ksprobs2} As in Fig.~\ref{fig:ksprobs1}, but for intervals of the kinematic age, $t_{\rm kin}$, and for the remaining sample of 33 pulsars, after excluding pulsars for which $t_{\rm kin}$ was ambiguous. For comparison, we have over-plotted the central values of $p_{\rm KS}(\tau_{\rm c})$ with light-grey dots, assuming the same scale for $\tau_{\rm c}$ as that of $t_{\rm kin}$ shown. The left-most data point, for $t_{\rm kin}<100$ kyr, shown with dashed lines, corresponds to the distribution of $\Psi$ values for the Crab and Vela pulsars, and PSRs J0538+2817 and J1952+3252, if these pulsars are included in our analysis with their true ages: $t_{\rm Crab}=958$ y; $t_{\rm Vela}\sim 25$ kyr; $t_{\rm J0538+2817}\sim 30$ kyr; and $t_{\rm J1952+3252}\sim 51$ kyr (see text for references).} 
\end{figure}

We have calculated the probability of rejecting uniformity, $p_{\rm KS}$, using only the aforementioned 5 pulsars, using the independent estimates of their true age. Apart from the age of PSR J1801$-$2451, for the rest of the pulsars these estimates are significantly younger than both $\tau_{\rm c}$ and $t_{\rm kin}$, which securely places them in the $0$--$100$ kyr interval. If we use the published $\Psi$ values for these pulsars, including that for PSR J1801$-$2451, then uniformity is rejected with only $p_{\rm KS}=76^{+17}_{-34}\%$ probability. However, as was stated earlier, the age of PSR J1801$-$2451 is quite uncertain and it could be significantly older than 100 Myr (Gaensler \& Frail 2000\nocite{gf00}). On that account, if we exclude this pulsar from the above sample, we obtain a much more significant rejection of uniformity, at the level of $94^{+5}_{-16}\%$. The overall trend of spin--velocity correlation with age is shown in Fig.~\ref{fig:ksprobs2}, where we have also added the corresponding $p_{\rm KS}$ value for the $0$--$100$ Myr interval from the 4 young pulsars. It can be seen from that plot that our general conclusion of a strong spin--velocity correlation for $\ll 10$ Myr that diminishes beyond 10 Myr is maintained.

\section{Discussion} 
\label{sec:Discussion} 

\subsection{Pulsar Distances} 
\label{subsec:puldists}
\begin{figure} 
\vspace*{10pt} 
\includegraphics[width=0.48\textwidth]{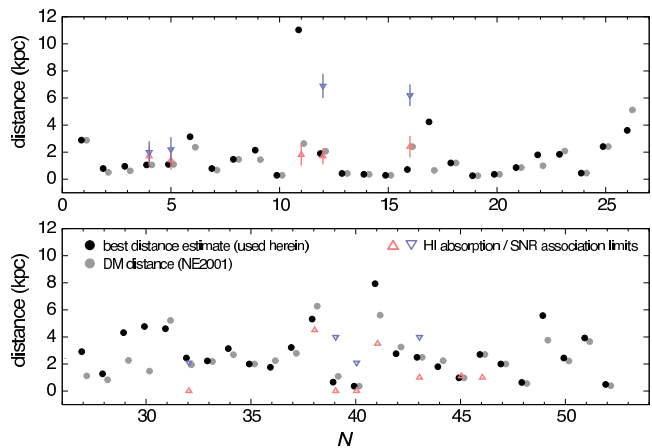} \caption{\label{fig:dmdistdiff} The best distance estimates of the 52 pulsars in our sample (black), contrasted with those based on the pulsar DM and the NE2001 model (gray), and the lower (red) and upper (blue) limits based on HI absorption or SNR associations. The index number $N$ on the horizontal axis corresponds to that shown in the first column of Table~\ref{tab:tab2}.} 
\end{figure}

In this section, we discuss another important factor that can potentially affect the above conclusions: the large uncertainties on pulsar distances. As was mentioned in the Introduction, DM-based estimates of pulsar distances are often significantly different from distances derived from SN associations, HI absorption or parallax measurements (e.g.~Chaterjee et al.~2009\nocite{cbv+09}; Deller et al.~2009\nocite{dtbr09}; Schnitzeler~2012\nocite{sch12}). Fig.~\ref{fig:dmdistdiff} shows a scatter plot of the best available distances for our sample of pulsars (black bullets), as well as alternative distance estimates based on the NE2001 electron-density model (grey bullets; \nocite{cl02}Cordes \& Lazio 2002); also, where available, lower and upper limits on the distance based on HI absorption and SNR associations are shown (triangles). It is clear from that plot that the differences between distances based on the model and those based on alternative methods can range from a few percent to several tens of percent; and in some cases the difference can be of the order of a few hundreds percent. It is hence important to examine the impact that a wrong estimate on the distance would have on the kinematic ages.

\begin{figure} 
\vspace*{10pt} 
\includegraphics[width=0.48\textwidth]{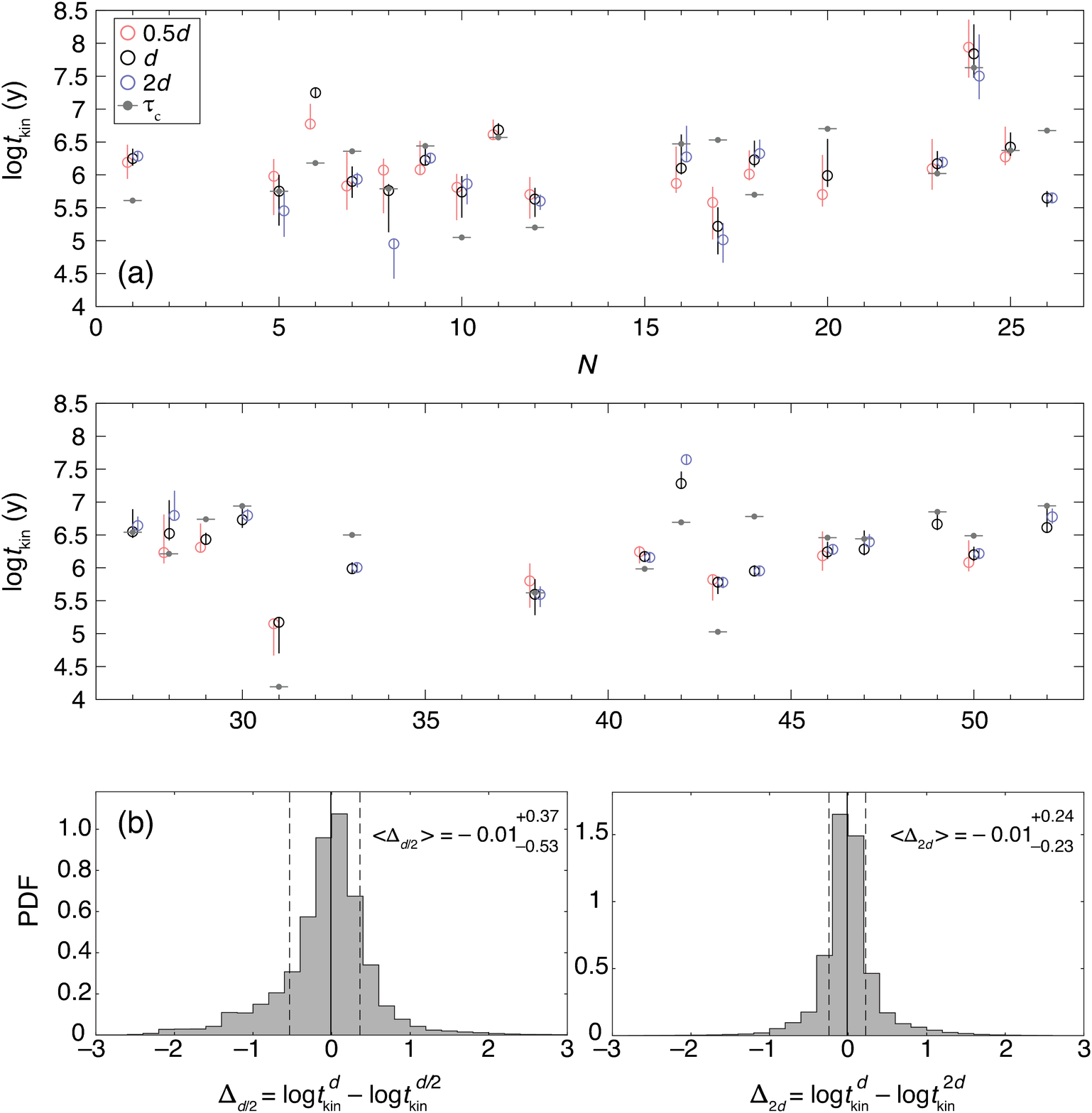} \caption{\label{fig:tkinDistPlot} (a) Top panels: Comparison of the kinematic ages of the 33 pulsars in our selected sample (black) with those derived using the same analysis but assuming the pulsars are at half their published distance (red) and at twice their published distance distance (blue). For the sake of comparison, the value of $\tau_{\rm c}$ for each pulsar is also shown as a grey dot with a horizontal bar. The index number $N$ on the horizontal axis corresponds to that shown in the first column of Table~\ref{tab:tab2}. Missing values in this plot imply that no solution was found within $\tau_1$ for the corresponding pulsar, using the corresponding distance. (b) Bottom panels: (left) PDF of the difference $\Delta_{d/2}$, between the $\log t_{\rm kin}$ values corresponding to distance $d$ and to $d/2$, for the selected 33 pulsars, and (right) of the difference $\Delta_{2d}$ between the $\log t_{\rm kin}$ values corresponding to distance $d$ and to $2d$. The value of the median of $\Delta_{d/2}$ and $\Delta_{2d}$ (black vertical lines) and their 68\% CLs (dashed vertical lines) are shown in the inset keys.} 
\end{figure}

We have reproduced the $t_{\rm kin}$ distributions for all pulsars in our sample assuming their distance is 0.5 and 2 times the published distance, $d$; the choice of varying the published distance by factor 2 was motivated by the worst-case discrepancies between the various distance estimates shown in Fig.~\ref{fig:dmdistdiff}. The resulting values of $t_{\rm kin}^{d/2}$, for pulsar distances equal to $0.5d$, and $t_{\rm kin}^{2d}$, for pulsar distances equal to $2d$, are shown in Table~\ref{tab:tab3}. A visual comparison between all three cases, i.e.~for pulsar distances equal to $d$, $0.5d$ and $2d$, is provided in the scatter plot of Fig.~\ref{fig:tkinDistPlot}a. In a few cases, our simulation did not converge to a solution within $\tau_1$, when the alternative distances were used: the data points corresponding to those cases are missing from the plot. Furthermore, as was done for Table~\ref{tab:tab2}, for pulsars with an ambiguous determination of $t_{\rm kin}$, we report only the kinematic age corresponding to the maximum of the PDF, and without quoting errors; the corresponding data points are also omitted from the scatter plot of Fig.~\ref{fig:tkinDistPlot}. For reference, the same plot shows the value of $\tau_{\rm c}$, for each pulsar.

At first glance, Fig.~\ref{fig:tkinDistPlot}a reveals that for the majority of pulsars both $t_{\rm kin}^{d/2}$ and $t_{\rm kin}^{2d}$ are consistent with the values of $t_{\rm kin}^d$, within the quoted errors. In order to assess this apparent similarity more quantitatively, we calculated the difference between $\log t_{\rm kin}^d$ and $\log t_{\rm kin}^{d/2}$, i.e.~$\Delta_{d/2}=\log t_{\rm kin}^d-\log t_{\rm kin}^{d/2}$ and between $\log t_{\rm kin}^d$ and $\log t_{\rm kin}^{2d}$, i.e.~$\Delta_{2d}=\log t_{\rm kin}^d-\log t_{\rm kin}^{2d}$, at every MC iteration of our simulation. The distributions of $\Delta_{d/2}$ and $\Delta_{2d}$, for all 33 pulsars are shown in Fig.~\ref{fig:tkinDistPlot}b. It is clear from these distributions that, on average, the modification to the pulsar distance does not significantly impact on $t_{\rm kin}$, considering the errors on the latter. More specifically, we find that $\langle\Delta_{d/2}\rangle=-0.01^{+0.37}_{-0.53}$, where the quoted 68\% interval is comparable to the standard deviation of the $p(t_{\rm kin})$ for the same sample (see Fig.~\ref{fig:ageDists}). On the other hand, the distribution of $\Delta_{2d}$ appears somewhat narrower, with $\langle\Delta_{2d}\rangle=-0.01^{+0.23}_{-0.24}$, revealing the lesser impact on $t_{\rm kin}$ by doubling the distance compared to that by halving it. 

Overall, we can conclude that the impact of distance errors on the determination of $t_{\rm kin}$, at least for our pulsar sample, can be neglected in comparison to the uncertainties arising from the unknown $v_r$ and $z_{\rm birth}$. There are 3 exceptions to this in our sample, PSRs J0452$-$1759, J0538+2817 and J1937+2544, for which the difference in $t_{\rm kin}$ is greater that $1\sigma$ between the different distance assumptions. The reason for the significant impact of the distance on the kinematic age for these 3 pulsars cannot be simply traced down to having special properties: neither their positional ($l$,$b$) nor their kinematic ($v_l$,$v_b$) parameters are outliers in our sample, which would account for the divide. Clearly, the impact of distance errors on estimates of $t_{\rm kin}$ warrants a complex description.

\begin{table*}
\renewcommand{\arraystretch}{1.3}
\caption{\label{tab:tab3}  
Kinematic ages of 33 pulsars with reliable estimates of $t_{\rm kin}$, assuming they are at different distances: (Column 4) kinematic age assuming the published pulsar distances, $d$ (shown in column 7 of Table~\ref{tab:tab2}), i.e.~$\log t_{\rm kin}^d$;  (Column 5) kinematic age assuming half the published distance, i.e.~$\log t_{\rm kin}^{d/2}$; and (Column 6) kinematic age assuming twice the published distance, i.e.~$\log t_{\rm kin}^{2d}$. Undeterminable solutions of $t_{\rm kin}$, for a particular distance assumption and pulsar are shown as dashes. Pulsars for which the $t_{\rm kin}$ determination is ambiguous for the particular distance assumption are accompanied only by the $t_{\rm kin}$ value corresponding to the PDF maximum. Also, for reference, Column 3 shows the pulsar characteristic age. A scatter plot of the tabulated data is shown in Fig.~\ref{fig:tkinDistPlot}.}
\normalsize
\centering
\begin{tabular}{@{}llrrrr}
\\
\hline
$N$ &    PSR                & $\log \tau_{\rm c}$ & $\log t_{\rm kin}^d$    & $\log t_{\rm kin}^{d/2}$  &  $\log t_{\rm kin}^{2d}$      \\
    &                       & [yr]                & [yr]                    &  [yr]                     &  [yr]                         \\
\hline
1   &    J0139+5814         &  5.6                &	$6.3_{-0.1}^{+0.1}$   & $6.2_{-0.3}^{+0.3}$   &   $6.3_{-0.1}^{+0.1}$    \\
5   &    J0358+5413         &  5.8                &	$5.7_{-0.5}^{+0.3}$   &	$6.0_{-0.6}^{+0.3}$   &   $5.5_{-0.4}^{+0.3}$    \\
6   &    J0452$-$1759       &  6.2                &	$7.3$                 &	$6.8_{-0.1}^{+0.3}$   &   --                     \\
7   &    J0454+5543         &  6.4                &	$5.9_{-0.2}^{+0.2}$   &	$5.8_{-0.4}^{+0.5}$   &   $5.9_{-0.1}^{+0.1}$    \\
8   &    J0538+2817         &  5.8                &	$5.8_{-0.6}^{+0.1}$   &	$6.1_{-0.7}^{+0.2}$   &   $5.0_{-0.5}$           \\
9   &    J0630$-$2834       &  6.4                &	$6.2_{-0.1}^{+0.2}$   &	$6.1_{-0.1}^{+0.4}$   &   $6.3_{-0.1}^{+0.1}$    \\
10  &    J0659+1414         &  5.0                &	$5.7_{-0.4}^{+0.3}$   &	$5.8_{-0.5}^{+0.2}$   &   $5.9_{-0.3}^{+0.4}$    \\
11  &    J0738$-$4042       &  6.6                &	$6.7^{+0.1}$          &	$6.6^{+0.2}$          &   --                     \\
12  &    J0742$-$2822       &  5.2                &	$5.6_{-0.3}^{+0.2}$   &	$5.7_{-0.4}^{+0.3}$   &   $5.6_{-0.1}^{+0.1}$    \\
16  &    J0837+0610         &  6.5                &	$6.1_{-0.1}^{+0.5}$   &	$5.9_{-0.1}^{+0.6}$   &   $6.3_{-0.1}^{+0.5}$    \\
17  &    J0837$-$4135       &  6.5                &	$5.3_{-0.5}^{+0.2}$   &	$5.9_{-0.7}^{+0.0}$   &   $5.0_{-0.3}^{+0.3}$    \\
18  &    J0922+0638         &  5.7                &	$6.2_{-0.1}^{+0.3}$   &	$6.0_{-0.1}^{+0.4}$   &   $6.3_{-0.1}^{+0.2}$    \\
20  &    J1136+1551         &  6.7                &	$6.0_{-0.2}^{+0.6}$   &	$5.7_{-0.2}^{+0.6}$   &   --                     \\
23  &    J1453$-$6413       &  6.0                &	$6.1_{-0.2}^{+0.2}$   &	$6.1_{-0.3}^{+0.5}$   &   $6.2_{-0.1}^{+0.1}$    \\
24  &    J1456$-$6843       &  7.6                &	$7.8_{-0.4}^{+0.4}$   &	$7.9_{-0.5}^{+0.4}$   &   $7.5_{-0.3}^{+0.6}$    \\
25  &    J1509+5531         &  6.4                &	$6.4_{-0.1}^{+0.2}$   &	$6.3_{-0.1}^{+0.5}$   &   --                     \\
26  &    J1604$-$4909       &  6.7                &	$5.6_{-0.1}^{+0.1}$   &	$7.7$ \ (!)           &   $5.7_{-0.1}^{+0.1}$    \\
27  &    J1645$-$0317       &  6.5                &	$6.6_{-0.1}^{+0.3}$   &	$6.3$ \ (!)           &   $6.6_{-0.1}^{+0.1}$    \\
28  &    J1709$-$1640       &  6.2                &	$6.5_{-0.1}^{+0.5}$   &	$6.2_{-0.2}^{+0.6}$   &   $7.4_{-0.6}^{+0.1}$    \\
29  &    J1735$-$0724       &  6.7                &	$6.4_{-0.1}^{+0.1}$   &	$6.3_{-0.1}^{+0.4}$   &   --                     \\
30  &    J1740+1311         &  6.9                &	$6.7_{-0.1}^{+0.3}$   &	$7.9$ \ (!)           &   $6.8_{-0.1}^{+0.1}$    \\
31  &    J1801$-$2451       &  4.2                &	$5.2_{-0.5}$          &	$5.2_{-0.5}$          &   --                     \\
33  &    J1844+1454         &  6.5                &	$6.0_{-0.1}^{+0.1}$   &	$6.0$ \ (!)           &   $6.0^{+0.1}$           \\
38  &    J1915+1009         &  5.6                &	$5.6_{-0.3}^{+0.2}$   &	$5.8_{-0.4}^{+0.3}$   &   $5.6_{-0.2}^{+0.1}$    \\
41  &    J1935+1616         &  6.0                &	$6.2_{-0.1}^{+0.1}$   &	$6.2_{-0.2}^{+0.1}$   &   $6.2$                  \\
42  &    J1937+2544         &  6.7                &	$7.3^{+0.3}$          &	$7.3$ \ (!)           &   $7.6$                  \\
43  &    J1952+3252         &  5.0                &	$5.8_{-0.2}^{+0.1}$   &	$5.8_{-0.3}^{+0.1}$   &   $5.8_{-0.1}^{+0.1}$    \\
44  &    J1955+5059         &  6.8                &	$5.9_{-0.1}^{+0.1}$   &	$7.7$ \ (!)           &   $6.0$                  \\
46  &    J2022+2854         &  6.5                &	$6.3_{-0.1}^{+0.2}$   &	$6.2_{-0.2}^{+0.4}$   &   $6.3_{-0.1}^{+0.1}$    \\
47  &    J2022+5154         &  6.4                &	$6.3_{-0.1}^{+0.3}$   &	$7.5$ \ (!)           &   $6.4_{-0.1}^{+0.1}$    \\
49  &    J2157+4017         &  6.8                &	$6.6_{-0.1}^{+0.2}$   &	$8.0$ \ (!)           &   --                     \\
50  &    J2219+4754         &  6.5                &	$6.2_{-0.1}^{+0.1}$   &	$6.1_{-0.1}^{+0.3}$   &   $6.2_{-0.1}^{+0.1}$    \\
52  &    J2305+3100         &  6.9                &	$6.6_{-0.1}^{+0.4}$   &	$7.8$ \ (!)           &   $6.8_{-0.1}^{+0.1}$    \\
\hline
\end{tabular}

\footnotesize \flushleft (!) Complex PDF with multiple significant peaks and in some cases discontinuities. Only the value corresponding to the maximum of the PDF is tabulated.  \\ \normalsize

\end{table*}

\subsection{Pulsar Braking Indices} 
\label{subsec:brakindx} 
\subsubsection{In general}
\begin{figure} 
\vspace*{10pt}
\includegraphics[width=0.48\textwidth]{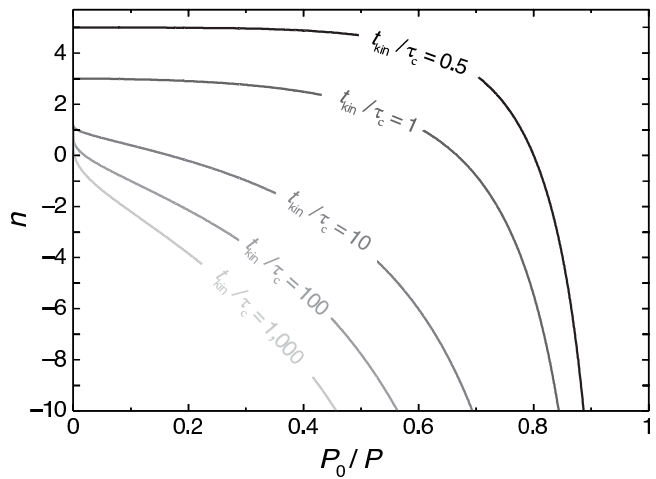}
\caption{\label{fig:bivp0} 
Pulsar braking index, $n$, as a function of birth period, $P_0$ (normalised by the present period, $P$), for a range of $t_{\rm kin}/\tau_{\rm c}$ values.}
\end{figure}

Despite being very difficult to measure, there are 8 pulsars so far with reliable braking indices in the literature (Lyne et al.~1993,1996\nocite{lps93}\nocite{lpgc96}; Middleditch et al.~2006\nocite{mmw+06}; Livingstone et al.~2007\nocite{lkg+07}; Weltevrede et al.~2011\nocite{wje11}; Espinoza et al.~2011\nocite{elk+11}). Although the pulsar sample is small, it is clear that none of them, in fact, have $n=3$, whereas the majority have $2<n<3$ with the rest of the values being $\sim 1$. Furthermore, if we assume that the PDFs of $t_{\rm kin}$ derived from our simulation correspond to those of true pulsar ages, we see that for a number of pulsars in our selected sample $\tau_{\rm c}$ falls several $\sigma$ away from the most likely value of $t_{\rm kin}$. The possible reasons for such significant discrepancies can be better understood if we examine the dependence of $n$ on $t_{\rm kin}$ and $P_0$, as is expressed by Eq.~\ref{eq:truage}: 
\begin{equation} 
\label{eq:ibrakDt2} 
n= 1+2\frac{\tau_{\rm c}}{t_{\rm kin}}\left[1-\left(\frac{P_0}{P}\right)^{n-1}\right]
\end{equation}
As $n$ is non-separable from $P_0$, in this equation, we can only solve it numerically for a range of $t_{\rm kin}/\tau_{\rm c}$ and $P_0/P$ values. Fig.~\ref{fig:bivp0} shows $n$ as a function of $P_0/P$, for different values of $t_{\rm kin}/\tau_{\rm c}$. The most evident feature of $n(P_0/P)$ in this plot is that $n$ approaches its maximum value of $1+2(\tau_{\rm c}/t_{\rm kin})$ as $P_0/P\rightarrow 0$, for any $t_{\rm kin}/\tau_{\rm c}$. 

Moreover, as was also discussed in Section~\ref{subsec:p0nbias}, it is clear that for a constant $n$, a larger pulsar age implies a smaller $P_0$ and vice versa: i.e.~a pulsar born slow-spinning would spin down to its present spin period, under a constant $n$, faster than the same pulsar born fast-spinning. Hence, pulsars like PSR J1955+5059, for which $t_{\rm kin}\sim 0.1\tau_{\rm c}$, may have been born with $P_0\approx 482$ ms and spun down with a constant $n=3$ or possibly with $P_0\ll P$ and spun down with $n=1+2\tau_{\rm c}/t_{\rm kin}\approx 17$, or perhaps, what is a more likely scenario, with another combination of $n$ and $P_0$ that lies between those values of $n$ and $P_0/P$, for $t_{\rm kin}/\tau_{\rm c}\sim 0.1$. Although, it should be noted that the value of $n$ for such small age fractions remains roughly constant and equal to its maximum allowed value for up to $P_0/P\sim 0.95$, as can already be inferred from Fig.~\ref{fig:bivp0}. Hence, unless we are willing to accept extreme values of $n$ for PSR J1955+5059, the results of our simulation suggest that it was very likely born with $P_0\sim P$. 

In contrast, pulsars with $t_{\rm kin}\gg \tau_{\rm c}$, like PSR J0922+0638, pose an interesting problem, as the discrepancy between $t_{\rm kin}$ and $\tau_{\rm c}$ cannot be reconciled under $n=3$, for any value of $P_0/P$ ({\em N.B.}~pure magnetic dipole braking only allows $t_{\rm kin}\leq \tau_{\rm c}$). For those cases, we are forced to accept that $n<3$, if the pulsar was born close to the GP. In our selected sample, there are 7 pulsars for which we can confidently exclude that $t_{\rm kin}<\tau_{\rm c}$ under the assumptions of our simulation. Critically, as we saw in Section~\ref{subsec:puldists}, an error on the assumed pulsar distance could modify $p(t_{\rm kin})$ to a certain degree. However, this was found to be a mitigating factor only in one case, that of PSR J0738$-$4042, whose $t_{\rm kin}$ could be made consistent with $\tau_{\rm c}$ if the pulsar is assumed to be 50\% closer to the Sun (see Table~\ref{tab:tab3}). For the rest of the pulsars, a modification to their distance by up to a factor of 2 cannot reconcile the fact that $p(t_{\rm kin})$ is inconsistent with pure magnetic dipole braking.

\subsubsection{Application to the selected sample}
The parametric curves of $n(P_0/P)$ in Fig.~\ref{fig:bivp0} give us a general idea of the expected distribution of $n$ for pulsars of different ages, born spinning slowly or spinning fast. In particular, for $P_0/P\approx 0$ we have $n\approx 1+2(\tau_{\rm c}/t_{\rm kin})\gtrsim 1$, with the upper limit restricted only by the minimum value of $t_{\rm kin}/\tau_{\rm c}$ --- e.g.~the lower $1\sigma$ CL on $t_{\rm kin}/\tau_{\rm c}$ for PSR J0837$-$4135, being $\approx 0.02$, would yield $n\approx 100$. On the other hand, for $P_0/P \neq 0$, the range of values for $n$ is unrestricted, and can even be negative: the lower bound on $n$ only comes from the largest $P_0/P$ we are willing to accept. In those cases, small values of $t_{\rm kin}/\tau_{\rm c}$ (e.g.~$<1$), as in the case of PSR J1955+5059, would yield $n\ll 1+2\tau_{\rm c}/t_{\rm kin}$ only for $P_0\sim P$, whereas pulsars with large $t_{\rm kin}/\tau_{\rm c}$, like PSR J0922+0638, would yield $n<0$ for all but the smallest values of $P_0/P$. 

\begin{figure} 
\vspace*{10pt}
\includegraphics[width=0.48\textwidth]{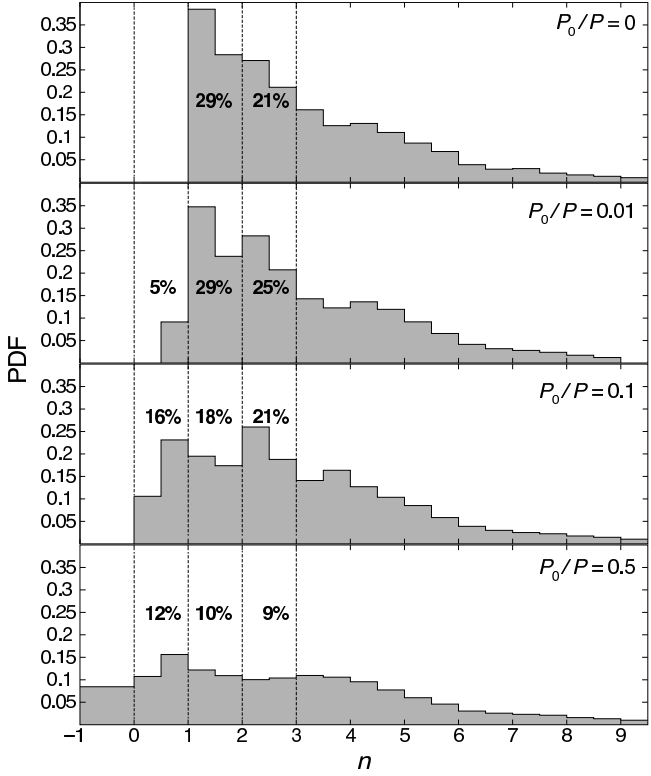}
\caption{\label{fig:nbrakDists} 
PDFs of the braking index, $n$, calculated from the PDFs of $t_{\rm kin}$ and Eq.~\ref{eq:ibrakDt2}, for the 33 selected pulsars of Table~\ref{tab:tab3}. The distributions shown correspond to (from top to bottom) $P_0/P=0, 0.01, 0.1$ and 0.5. The vertical dotted lines demarcate the intervals between $n=0,1,2$ and 3, where we also show the percentage of values that fall into each interval.}
\end{figure}

One of the interesting open questions regarding pulsar evolution is the dominant spin-down mechanism responsible for the bulk movement of pulsars on the $P$--$\dot{P}$ diagram. The small number of measured braking indices does not allow for a statistical analysis. However, our simulation allows us to calculate the distribution of $n$ from $p(t_{\rm kin})$, for each of the selected 33 pulsars, using Eq.~\ref{eq:ibrakDt2} and assuming a value for $P_0/P$. The aggregate distribution of $n$ for all 33 pulsars assuming $P_0/P=0, 0.01, 0.1$ and 0.5 is shown if Fig.~\ref{fig:nbrakDists}. As expected, assuming that all our pulsars were born spinning infinitely fast ($P_0/P=0$) results in all values of $n$ being distributed above 1, with a noticeable peak near $n=1$. The reason for the accumulation of values in the bottom bin of that distribution is that all solutions with $t_{\rm kin}\gg\tau_2$ will be binned near $n=1$, since slower spin-down is not allowed. In other words, the distribution of $n$ for $P_0/P=0$ is populated with values that are in the majority upper limits to the true values of the braking index. If we allow finite values of $P_0$, it can be seen in the rest of the distributions that a significant fraction of solutions previously corresponding to $n\approx 1$, is now distributed between $n=0$ and $n=1$. Finally, in the more extreme case of $P_0/P=0.5$, the PDF of $n$ is practically uniform across a wide range of $n$ and even allows for $n<0$.

In conclusion, based on the distributions of $n$ for our sample, under different assumptions for $P_0/P$, it is difficult to determine the most likely range of braking indices, independently of $P_0$. Indeed, if we assume that all 33 pulsars were born with $P_0/P\lesssim 0.01$, then $n\sim 1$--2 seems the most likely. However, we cannot exclude the possibility of pulsars being born with spin periods much nearer their measured value; in which case, as can be seen in Fig.~\ref{fig:nbrakDists}, spin-down models with $0<n<3$ appear to be an equally likely scenario.

\begin{figure} 
\vspace*{10pt} 
\includegraphics[width=0.48\textwidth]{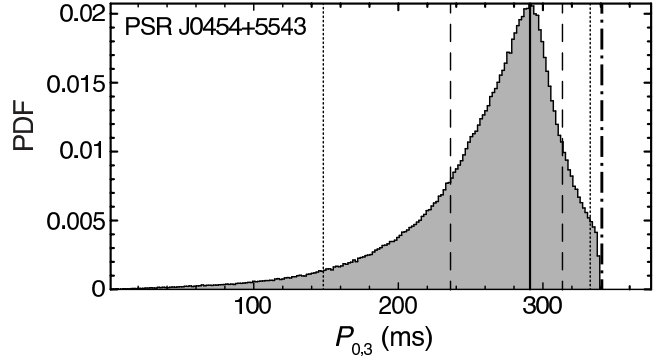} \caption{\label{fig:agep0dist} The birth-period PDF, $p(P_{0,3})$, calculated from the kinematic-age distribution of PSR J0454+5543, $p(t_{\rm kin})$, assuming $n=3$ (pure magnetic-dipole braking). The solid line corresponds to the peak of the PDF and the dashed and dotted lines, to the 68\% and 95\% confidence limits, respectively. The rightmost, dash-dotted line marks the presently measured period for this pulsar.} \end{figure}

\subsection{Pulsar Birth Periods}
\label{sec:birthpers}

\subsubsection{In general}
 
\begin{figure} 
\vspace*{10pt}
\includegraphics[width=0.48\textwidth]{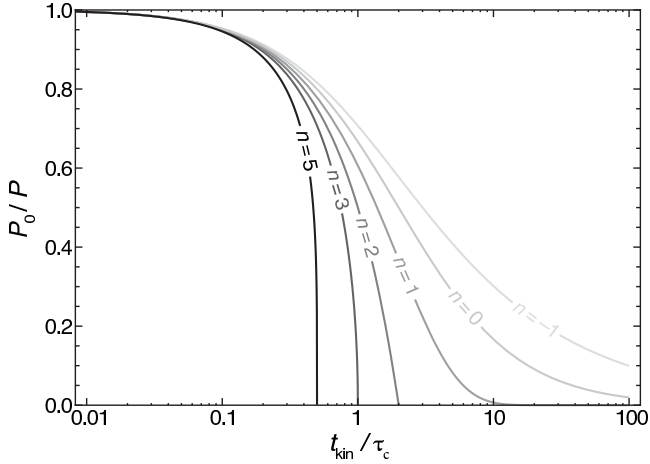}
\caption{\label{fig:p0vtkin} 
Pulsar spin period at birth, $P_0$ (normalised by the present period, $P$) as a function of the kinematic age, $t_{\rm kin}$ (normalised by $\tau_{\rm c}$), for a range of braking index values.}
\end{figure}

In addition to the determination of $n$ from $t_{\rm kin}$, assuming different values for $P_0/P$, one can invert the problem and determine $P_0$ for different spin-down models, assuming a value for the braking index, $n$: i.e.~by setting $t_{\rm true}=t_{\rm kin}$ and solving Eq.~\ref{eq:truage} for $P_0$, 
\begin{equation}
\label{eq:birthperiod}
P_0=P\left[1-\frac{(n-1)t_{\rm kin}}{2\tau_{\rm c}}\right]^{\frac{1}{n-1}}
\end{equation}
where $P$ is the observed pulsar period. Note that the maximum allowed value of $t_{\rm kin}$ for a given $n$, for which $P_0$ remains a real number in this equation is $\tau_n=2\tau_{\rm c}/R(n-1)$; here, $R(n-1)$ is the ramp function, which is equal to $n-1$ for $n\geq 1$ and 0 everywhere else. This is the generalised form of the characteristic age for different $n$, which assumes that $P_0/P=0$. The reader will notice that, according to the above definition of $\tau_n$, the characteristic age for $n=1$ is infinity, which is true if $P_0=0$ is assumed; but $\tau_1$ is finite as we have defined it in Eq.~\ref{eq:charage1}. The allowed range of $t_{\rm kin}/\tau_{\rm c}$ for different spin-down models is schematically shown in Fig.~\ref{fig:p0vtkin}. 

Assuming a pure magnetic-dipole spin-down, i.e.~$n=3$, Eq.~\ref{eq:birthperiod} becomes
\begin{equation}
\label{eq:birthperiod_n3}
P_{0,3}=P\sqrt{1-\frac{t_{\rm kin}}{\tau_{\rm c}}}
\end{equation}
In this section, when appropriate, we use two subscript indices for $P$: the first index denotes the time at which the period is calculated (e.g.~at birth, $t=0$), and the second element, the assumed braking index in that calculation (e.g.~$n=3$). 


We have used Eq.~\ref{eq:birthperiod_n3} to calculate the distributions of $P_{0,3}$ for each of the 27 from the 33 selected pulsars, for which $p_3=p(t_{\rm kin}\leq \tau_{\rm c})>0$. Fig.~\ref{fig:agep0dist} shows the distribution of $P_{0,3}$ for PSR J0454+5543. The value of $P_0$ corresponding to the maximum of the PDF for each of the 27 pulsars is shown in column 15 of Table~\ref{tab:tab2}.


In several cases, the calculation of the PDF of $P_{0,3}$, from Eq.~\ref{eq:birthperiod_n3}, excludes a significant fraction of $p(t_{\rm kin})$. An example that highlights this fact is that of PSR J1453+6413, for which $P_{0,3}$ can be calculated from only 15\% of the values produced in our simulation. Similarly, for $n=2$, $P_{0,2}$ can be calculated for about 74\% of those values; and, finally, $P_{0,2}$ can be calculated for all values of $t_{\rm kin}$. The shape of these PDFs depends directly on $n$ (Eq.~\ref{eq:birthperiod}), but since different choices of $n$ also allow different ranges of $t_{\rm kin}$, $n$ also influences the PDFs indirectly:

\begin{itemize}

\item[(a)] The direct dependence on $n$ can be estimated by calculating the first derivative of Eq.~\ref{eq:birthperiod} with respect to $n$. The fractional change of $P_0$, $\delta P_0/P_0$, as a function of the change in $n$, $\delta n$, is then 

\begin{equation}
\label{eq:birthperiodDn}
\frac{\delta P_0}{P_0}\approx\begin{cases} \frac{1}{2(1-n)}\left[\left(\frac{P_0}{P}\right)^{1-n}\left(\frac{t_{\rm kin}}{\tau_{\rm c}}\right)+\ln\left(\frac{P_0}{P}\right)^2\right]\delta n \\
\\
-\frac{1}{8}\left(\frac{t_{\rm kin}}{\tau_{\rm c}}\right)^2 \delta n  \ \ \ \ \ \ {\rm if} \  n\approx 1
\end{cases}
\end{equation}
It can be shown that $\delta P_0/P_0$ always has the opposite sign to $\delta n$. 
Therefore, assuming a smaller value of $n$ results in larger values of $P_0$ for the same $t_{\rm kin}$. \\

\item[(b)] Further to the above, if we assume a small change of the braking index by $\delta n \ll n$, then the range of $t_{\rm kin}$ that contributes to $p(P_0)$ will change by 
\begin{align}
\nonumber
\frac{\delta t_{\rm kin}}{\tau_{\rm c}} \equiv \frac{\delta \tau_n}{\tau_{\rm c}}=\frac{2}{R(n+\delta n -1)}-&\frac{2}{R(n-1)}\approx \\
&\approx -\frac{1}{2}\left(\frac{\tau_n}{\tau_{\rm c}}\right)^2\delta n 
\end{align}
So, an increase in $n$ shrinks the allowed range of $t_{\rm kin}$ and a decrease, expands it.
The effect of $\delta n$ on the distribution of $P_0$ depends on the value of $t_{\rm kin}$ relative to $\tau_m$, where $m=\max{(n,n+\delta n)}$. In general, the distribution of $P_0$ will be modified according to the following:
\begin{equation}
\label{eq:birthperiodDt}
\nonumber 
p[P_{0,n+\delta n}(t_{\rm kin})]=\begin{cases} \frac{p(\leq \tau_n)}{p(\leq \tau_{n+\delta n})}p[P_{0,n}(t_{\rm kin})] \ \ \ {\rm if} \  t_{\rm kin}\leq \tau_m \\ 
\\
H(\tau_{n+\delta n}-t_{\rm kin})p[P_{0,n+\delta n}(t_{\rm kin})]  \\
\ \ \ \ \ \ \ \ \ \ \ \ \ \ \ \ \ \ \ \ \ \ \ \ \ \ \ \ \ \ \ \ \ \ {\rm if} \  t_{\rm kin}>\tau_m\end{cases}
\end{equation}
where $p(\leq \tau_n)=\int_0^{\tau_n} p(t_{\rm kin})d t_{\rm kin}$. Although it is possible, as was shown earlier, to express the change in $P_0$ for $t_{\rm kin}\leq\tau_m$ analytically, the form of $p(P_0)$ for $t_{\rm kin}>\tau_m$ depends entirely on $p(t_{\rm kin})$ and is therefore unique to every pulsar. 

\end{itemize}

\begin{figure} 
\vspace*{10pt}
\includegraphics[width=0.48\textwidth]{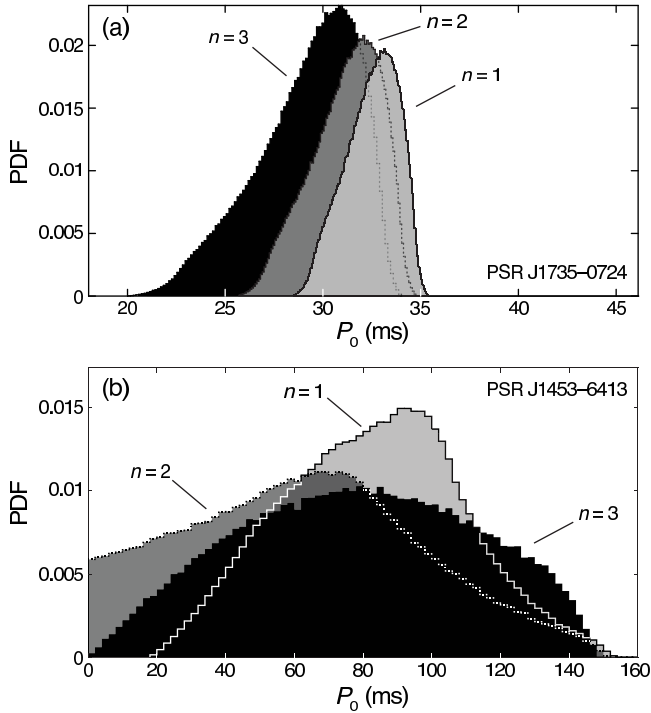}
\caption{\label{fig:j1453n123} 
Birth-period PDFs for (a) PSR J1735$-$0724 and (b) PSR J1453$-$6413, for different values of the braking index, i.e.~$n=1,2$ and 3, as indicated.}
\end{figure}

Bearing the above in mind, in Fig.~\ref{fig:j1453n123} we show two examples that highlight the direct and indirect effect of the choice of $n$ on the distribution of $P_0$. In those examples, we show the distributions $p(P_{0,1})$, $p(P_{0,2})$ and $p(P_{0,3})$ for PSRs J1735$-$0724 and J1453$-$6413, as they were derived from our analysis assuming 3 different values for $n$. Firstly, in Fig.~\ref{fig:j1453n123}a it can be seen that the PDF of $P_0$ is progressively shifted  to larger values of $P_0$ with decreasing $n$; this is also the case for the most probable values of $P_0$ in those distributions. This can be understood by looking at the $p(t_{\rm kin})$ of this pulsar in Fig.~\ref{fig:selectPDFs}, which lies entirely below $t_{\rm kin}=\tau_{\rm c}$; hence, in this case, the effect of $\delta n$ is solely due to the direct change of $P_0$ as is predicted by Eq.~\ref{eq:birthperiodDn}. On the other hand, Fig.~\ref{fig:j1453n123}b shows that for PSR J1453$-$6413 the different choices of $n$ yield distributions with very different shapes. Also, the most probable values of $P_0$ in these distributions do not change monotonically with $n$. This can be explained by the fact that the 3 different choices of $n$ sample significantly different fractions of the $p(t_{\rm kin})$ of this pulsar (Fig.~\ref{fig:selectPDFs}). Therefore, the shape of the PDFs of $P_0$ is determined both by the direct change of $P_0$ as a function of $n$ {\em and} by the indirect influence of $n$, as is described by Eq.~\ref{eq:birthperiodDt}.


\subsubsection{Application to the selected sample}
\begin{figure} 
\vspace*{10pt} \includegraphics[width=0.48\textwidth]{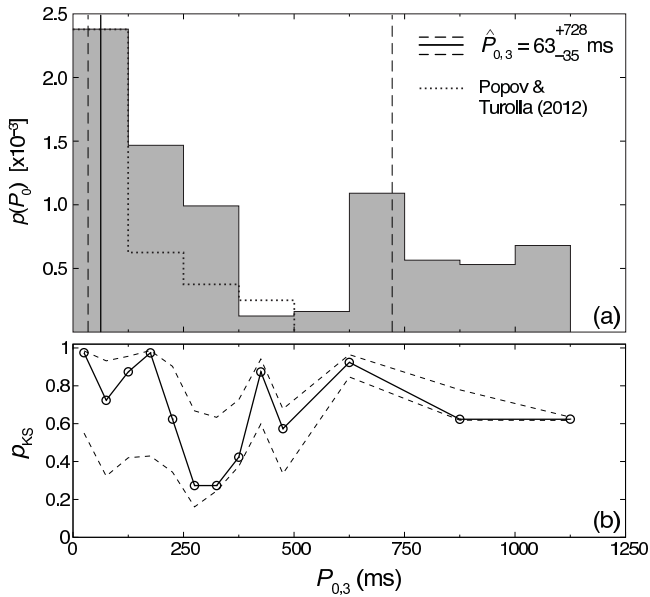} \caption{\label{fig:ksprobs3} (a) Birth-period distribution, $p(P_0)$, derived from a large number of simulated values of $t_{\rm kin}$ and Eq.~\ref{eq:birthperiod_n3}. In this plot, the solid line corresponds to the most-probable value of $P_0$, and the dashed lines, to the 68\% confidence limits around that value, respectively. In addition, the distribution of 30 birth periods of pulsars associated with SNRs, from the recent work of Popov \& Turolla (2012) is shown with a dotted line. (b) Dependence of the probability of rejecting the uniform $\Psi$ distribution, under the KS test, on the pulsar birth period, $P_0$, calculated for $n=3$. The plot details are given in Section~\ref{sec:birthpers}. Both plots are based on the selected sample of 27 pulsars, after excluding those with ambiguous kinematic-age determinations and for which the calculation of $P_0$ was possible under the assumption of $n=3$.} 
\end{figure}

In the previous paragraphs, we highlighted the fact that the shape of $p(P_{0,n})$ is sensitive to the choice of $n$. Keeping this in mind, we examined the aggregate distribution of the birth period by binning all $P_{0,3}$ values of the 27 pulsars, as we did for $n$ in the previous section. The normalised PDF, $p(P_{0,3})$, is shown in Fig.~\ref{fig:ksprobs3}a.

The derived distribution of the birth periods of 27 non-recycled pulsars from our work can be compared with the corresponding distribution that was published recently by Popov \& Turolla (2012)\nocite{pt12}: in that work, the authors inferred the birth periods of 30 pulsars based on their associated-SNR ages. As in our case, their work assumed standard magnetic-dipole braking ($n=3$) and concluded that a Gaussian distribution with a mean and standard deviation of 100 ms --- truncated at $P_0=0$ --- is a good description of the derived $P_0$ values. Their distribution is shown with a dotted line in Fig.~\ref{fig:ksprobs3}a. The most apparent difference between the two distributions is the much broader range of $P_0$ predicted by our simulation as opposed to the narrow distribution of Popov \& Turolla. This dissimilarity is confirmed by the two-sample KS test, which yields $\sim 0$ probability of these distributions coming from the same underlying distribution. Also, apart from the main peak at $P_0\sim 60$ ms, the derived distribution from our simulations exhibits a second long-period component for $P_0>600$ ms. The latter is mainly due to the calculated values for PSRs J0630$-$2834, J0837$-$4135, J0837+0610, J1136+1551, J1740+1311 and J2157+4017, which have $P\gtrsim 700$ ms and for which the probability of having $t_{\rm kin}\ll \tau_{\rm c}$ is high (see Table~\ref{tab:tab2}). Critically, these 6 pulsars are old and without SNR associations, and so they were not included in the sample of Popov and Turolla.

A number of SN-kick models predict a connection between $P_0$ and $\Psi$ (see e.g.~Ng \& Romani 2007\nocite{nr07}). According to these population-synthesis models, pulsars born with $P_0\lesssim20$ ms and those born with $P_0\gtrsim100$ ms tend towards alignment. For the short-period pulsars, the fast spin of the proto--neutron star tends to average out the kick directions, with the net kick being along the spin axis. Conversely, for the long-period pulsars, the proto--neutron star must have been spinning slowly at the start and would have been kicked radially along its spin axis, as to not have been spun up and acquire an orthogonal $\boldsymbol{v}_{\rm birth}$. According to the above simulations, pulsars with intermediate birth periods of 40--60 ms are expected to have $\Psi\gg0^\circ$: this is because these pulsars were spinning slow enough that the bulk of the initial kick was delivered at large angles, almost perpendicularly to the spin direction.

Our simulation gives us the opportunity to test the above predictions, using the $P_{0,3}$ distributions of the 27 pulsars. We followed the same procedure that was used in Figs.~\ref{fig:ksprobs1} and \ref{fig:ksprobs2}, wherein we calculated the corresponding KS probabilities of rejecting the uniform $\Psi$ distribution for a number of $P_0$ bins. In order to roughly match the scale over which changes in spin--velocity alignment are predicted according to the SN-kick models, we binned the values of $P_0$ into intervals of 50 ms, for $P_0<500$ ms, and into intervals of 250 ms, for the rest of the $P_0$ range: the upper limit for the finer binning was chosen to lie roughly on the border between the short-period and long-period components of $p(P_{0,3})$ (Fig.~\ref{fig:ksprobs3}a). The $1\sigma$ confidence intervals of the $P_{0,3}$ distributions for each pulsar are significant, i.e.~comparable to the size of the selected bins. And so, instead of assigning the most probable value of $P_0$ to each pulsar from its corresponding distribution, we randomly selected a large number of $P_0$ values from each pulsar's distribution: for each randomly selected set of 27 $P_0$ values, we calculated $p_{\rm KS}$ only for those pulsars whose $P_0$ value fell into the investigated $P_0$ bin. The final value of $p_{\rm KS}$ assigned to each bin corresponded to the most-probable value in the distribution of all $P_0$ values in the bin considered. Also, the 68\% CLs were calculated around the most-probable value. In this way, we incorporated $p(P_{0,3})$ into the final value of $p_{\rm KS}$. Fig.~\ref{fig:ksprobs3}b shows a scatter plot of the values of $p_{\rm KS}$ across the binned $P_0$ range. It is clear from that plot that the small number of available pulsars results in only a few pulsars being considered per MC iteration, in each bin. Consequently, the 68\% confidence intervals dominate over any possible dependence of $p_{\rm KS}$ on $P_0$. 

In summary, our efforts to extract information on pulsar birth periods based on kinematic simulations and an assumption of their spin-down were generally plagued by low statistics. Any tests against model predictions demand a relatively high resolution in $P_0$ space (i.e.~$\delta P_0\sim 20$ ms), which our limited sample could not fulfil --- allowing in the vast majority of cases for only 4 pulsars to be considered per 50-ms-wide $P_0$ bin. As in the case of pulsar ages, only a larger sample of suitable pulsars would help overcome these difficulties.

\section{Summary \& Conclusions}
\label{sec:conclusions}
\begin{figure} 
\vspace*{10pt}
\includegraphics[width=0.48\textwidth]{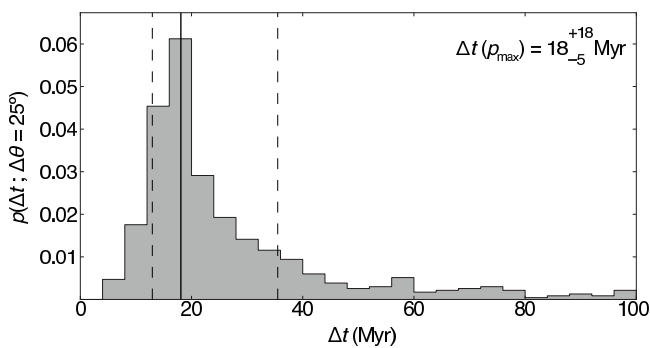}
\caption{\label{fig:dynagedist} 
Distribution of the minimum time interval, $\Delta t$, between the current position of the pulsars of this paper and their past position when their 3D velocity direction differed by $25^\circ$ from its present direction: i.e.~$\Delta\theta=25^\circ$. The value of $\Delta t$ corresponding to the peak of the distribution (solid line) and the 68\% CLs (dashed lines) are shown in the top right corner of the figure.}
\end{figure}

\begin{figure*} 
\vspace*{10pt}
\includegraphics[width=1.0\textwidth]{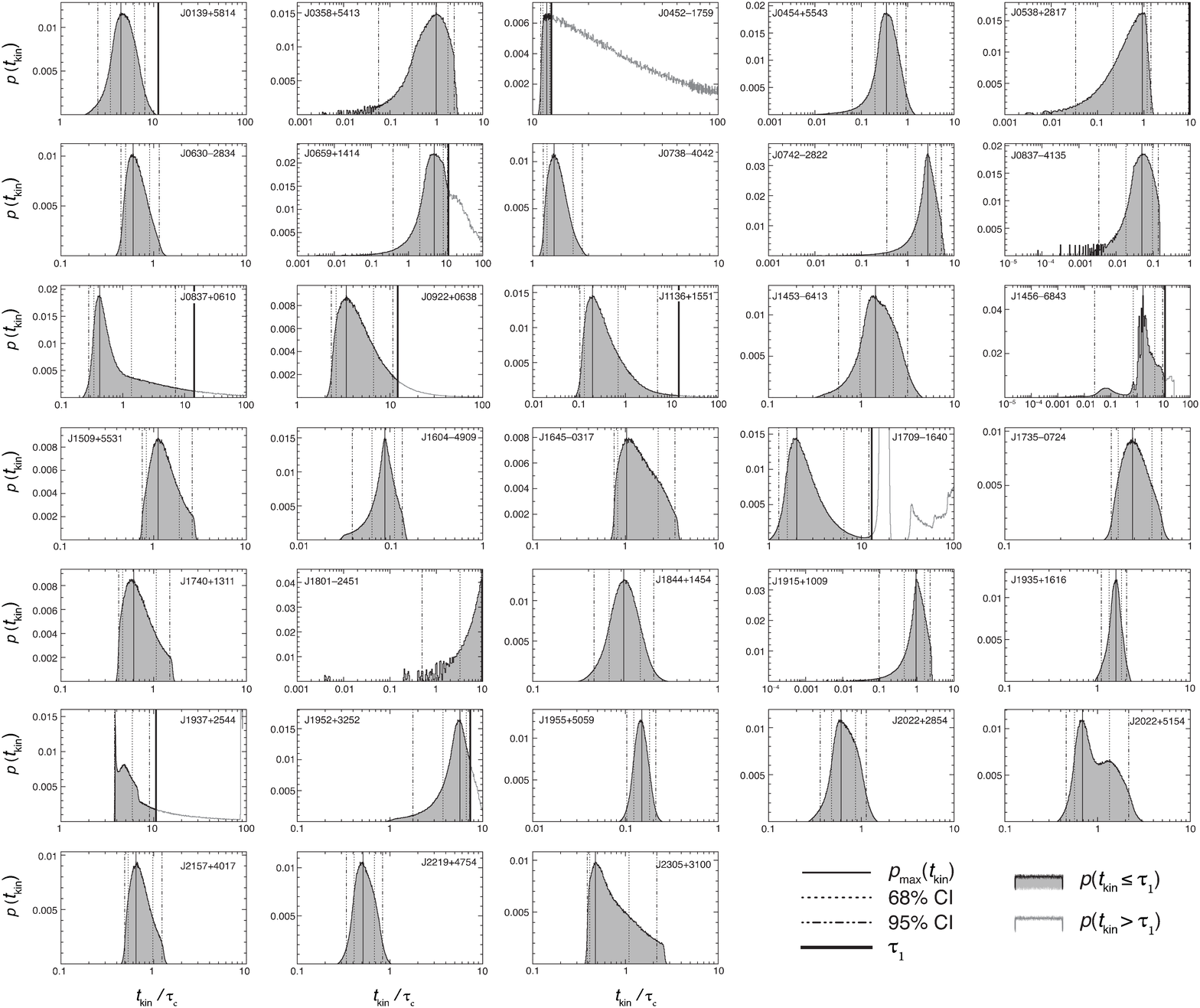}
\caption{\label{fig:selectPDFs} The $t_{\rm kin}$ PDFs of the selected 33 pulsars that were chosen for the spin--velocity alignment analysis.}
\end{figure*}

In this paper we investigated whether the high degree of correlation between the spin and velocity axes of pulsars depends on pulsar age. In the first part of our study, we used the characteristic ages ($\tau_{\rm c}$) of 58 non-recycled pulsars and showed that the strong correlation does not vanish for pulsars with characteristic ages above $\sim 10$ Myr, while we would expect any nascent correlation to have been washed out due to movement through the Galactic potential. This expectation is justified by the distribution of time intervals, $\Delta t$, during which the velocity direction of pulsars is significantly modified due to the Galactic gravitational potential. The distribution of $\Delta t$ was calculated for the pulsars in this paper, for a change in their velocity direction by $\Delta \theta = 25^\circ$ due to the galactic potential of Paczynski (see Fig.~\ref{fig:dynagedist}). The time interval corresponding to the peak of the distribution is $\Delta t=18_{-5}^{+18}$ Myr, which is close to an estimate of the dynamical time for the Milky Way (see Introduction). 

The conclusion of the first part of this work was based on pulsar characteristic ages, which are often unreliable: these are derived assuming that all pulsars are born spinning infinitely fast and that they spin down under a constant braking index of 3, by emitting pure magnetic-dipole radiation. To reach a more conclusive statement about the relationship between spin--velocity alignment and pulsar age, in the next part we considered an alternative, more accurate measure of a pulsar's age, based on the observed position, distance and proper motion of the pulsar: the kinematic age, $t_{\rm kin}$. The latter is defined as the travel time between the location of the pulsar at birth and that at present, through the Galactic gravitational potential. We have calculated the most-likely values of $t_{\rm kin}$ for 52 of the pulsars in our sample, under the reasonable assumption that they were born somewhere within 100 pc of the Galactic mid-plane (e.g.~Cordes \& Chernoff 1998\nocite{cc98}; Arzoumanian, Chernoff \& Cordes 2002\nocite{acc02}) and that they have travelled to the currently observed position through the gravitational potential of Paczynski (1990). The unknown radial velocities and birth heights were assumed to follow the distributions of Hobbs et al.~(2005) and Reed (2000), respectively, which have been derived from a large sample of pulsar proper motions and photographic magnitudes of OB stars. For each pulsar, our simulation was limited to a maximum age that we defined as the amount of time required for the pulsar to spin down to its present period, assuming it was born with a period of 1 ms and spins down with a braking index of 1: for the vast majority of pulsars in our sample, this limit was at least an order of magnitude older than their characteristic age. Our simulation resulted in 52 pulsars producing at least one intersection with the Galactic plane within the explored time interval. This is the largest sample of pulsar ages estimated based on kinematics to date.

Following the $t_{\rm kin}$ estimates, we tested the degree of spin--velocity alignment again, this time as a function of the more accurate kinematic age. To eliminate systematic uncertainties, caused by ambiguous age determinations due to multiple intersections with the Galactic plane, we selected only the 33 most reliable $t_{\rm kin}$: these were selected based on the complexity of the derived probability density functions of $t_{\rm kin}$. Unlike the case when $\tau_{\rm c}$ was used as the pulsar age, sorting the selected pulsars based on intervals in $t_{\rm kin}$ resulted in a diminishing degree of spin--velocity correlation with increasing age. Pulsars with ages greater than $\sim 10$ Myr, whose 3D velocities have almost certainly been significantly altered by the gravitational pull of the Galaxy, yielded no correlation. 
Hence, the present work --- together with our preceding study of spin--velocity alignment --- suggests that the orientations of the spin and velocity axes in young pulsars are correlated, while this is no longer observable for older pulsars due to the effect of the Galactic gravitational potential. 

Our results are tantalising, but must be strengthened with more data: e.g.~the interval corresponding to the oldest kinematic ages in our sample contained only 3 pulsars. Proper motion measurements of old, non-recycled pulsars are necessary to improve statistics in that interval. In addition, the spin-axis directions of these pulsars can be determined through polarisation measurements. Currently, a number of promising VLBI campaigns, which will result in hundreds of new pulsar proper motions, are in operation or being proposed (PSR$\pi$, Deller et al.~2011\nocite{dbc+11}; $e\Pi$, Vlemmings et al., $e$Merlin  Legacy Programme\nocite{vlemweb}). For many of these pulsars, these campaigns will also provide accurate distance measurements through parallax measurements, independent of electron-density modelling. Our work has shown that, for the vast majority of pulsars in our sample, a change in the distance by as much as a factor 2 (typical worst-case scenario for NE2001) does not affect the kinematic-age estimates, which remain consistent within $1\sigma$. However, for many pulsars the error on their distance derived from their DM can be larger still. Our simulations have shown that the effect of distance errors on the kinematic age in our simulations is complex --- and varies significantly with pulsar position and velocity --- so the availability of accurate distances will help eliminate such systematic uncertainties.
 
Upon the completion of the above VLBI campaigns, we will be able to combine new and existing measurements and improve the statistics. This will help answer many open questions about the nature of spin--velocity alignment, like the preference of SN kicks towards alignment or orthogonality; it will also allow for a distinction between the degree of alignment for pulsars born spinning fast and those born spinning slow. Finally, with a larger data set we can explore finer age intervals, which will provide a better handle on the characteristic time scale of the observability of spin--velocity alignment.  

Our work has shown that it is possible to systematically estimate pulsar ages independently of the choice of a spin-down model, while taking into account the kinematic properties of individual pulsars. In this work, we used this method to study the evolution of pulsar spin--velocity alignment with age, which required a pulsar sample with available absolute-polarisation information. Nevertheless, this method can be used independently with the hundreds of available pulsar proper motions to re-evaluate the ages of pulsars; this is invaluable information that can be used, in future, to study, amongst other things, the evolution of pulsars on the $P$--$\dot{P}$ diagram (Fauchere-Gigu\`ere \& Kaspi 2006\nocite{fk06}; Keane \& Kramer 2008\nocite{kk08}).
In this paper, we performed a preliminary statistical interpretation of the distributions of the birth periods and braking indices for our pulsars, based on the ratios between their characteristic and kinematic ages. In particular, we translated the distributions of kinematic age into distributions of braking index, under different assumptions for the value of the birth period. Under the usual assumption of $P_0/P=0$, we find that there is a clear preference towards $n=1$: however, this is a biased result caused by the restriction to $n\geq 1$ under this assumption. Allowing our pulsars to be born spinning slower results in distributions of $n$ that are noticeably more uniform, thus eliminating any preference towards a particular spin-down model. Inversely, assuming pure dipole magnetic braking, we calculated the birth-period distributions of 27 of the selected 33 pulsars from their $t_{\rm kin}$ distributions. The aggregate distribution of $P_0$ from this work is significantly wider compared to the birth-period distribution of SNR-associated pulsars. This can be explained by the very long birth periods ($P_0\sim 1$ s; comparable to presently observed periods) of several pulsars for which our simulation yields $t_{\rm kin}\ll \tau_{\rm c}$. 

In the near future, a large number pulsars will be discovered in all-sky pulsar surveys, like those currently underway with the Low Frequency Array (LOFAR; Stappers et al.~2011\nocite{sha+11}), and the Parkes and Effelsberg radio-telescopes (Keith et al.~2010\nocite{kjv+10}; Ng 2011\nocite{ng11}). Many of these pulsars will be discovered at low observing frequencies ($\sim 100$ MHz) and will thus be nearby because of signal-scattering limitations imposed by the interstellar medium over long distances (van Leeuwen \& Stappers 2010\nocite{ls10}). Hence, it will be possible to obtain parallax measurements, as was mentioned earlier, and estimate their proper motions and distances. Also, many more pulsars will be discovered at high latitudes, where scattering is less detrimental. The kinematic ages of those pulsars will be more reliably determined by methods such as that presented in our work. It will then be possible to conduct a study of the distributions of pulsar braking indices using large-number statistics, which was not available at the time of this work. Crucially, follow-up polarisation measurements of all the discovered pulsars will yield information on their spin-axis directions. The aggregate sample of hundreds of spin and velocity directions of known pulsars and those discovered in the above surveys will provide an unprecedented handle on the intriguing phenomenon of pulsar spin--velocity alignment.

Ultimately, the planned Square Kilometer Array (SKA) will discover tens of thousands of Galactic pulsars and will have the capability to directly measure proper motions for several thousands of them. This will help us address many open questions regarding the kinematic properties of pulsars, like which is the birth-velocity distribution of pulsars and what is the probability of spin--velocity alignment during a core-collapse supernova kick, for different pulsar properties at birth. 

\section{Acknowledgements}
The authors would like to thank Drs.~Norbert Wex and KJ~Lee for providing the {\em Runge-Kutta} integration code for our simulations and for contributing invaluable comments during the preparation of this document.

\newpage

\pagebreak

\bibliography{journals,modrefs,psrrefs,crossrefs}

\vfill

\end{document}